# How Thick is the Air-Water Interface? - A Direct Experimental Measurement of the Decay Length of the Interfacial Structural Anisotropy

Alexander P. Fellows[1], Álvaro Díaz Duque[1], Vasileios Balos[2], Louis Lehmann[3], Roland R. Netz[3], Martin Wolf[1], and Martin Thämer[1]*

[1] Fritz-Haber-Institut der Max-Planck-Gesellschaft, Faradayweg 4-6, 14195, Berlin, Germany

[2] Instituto Madrileño de Estudios Avanzados en Nanociencia (IMDEA Nanociencia), 28049, Madrid, Spain

[3] Department of Physics, Freie Universität Berlin, Arnimallee 14, 14195, Berlin, Germany

* Corresponding author

thaemer@fhi-berlin.mpg.de

(tel.): +49 (0)30 8413 5220

## Abstract

The air-water interface is a highly prevalent phase boundary with a far-reaching impact on natural and industrial processes. Water molecules behave differently at the interface compared to the bulk, exhibiting anisotropic orientational distributions, reduced intermolecular connectivity in the hydrogen bond network, and significantly slower dynamics. Despite many decades of research, the thickness of the structural anisotropy in the interfacial layer remains controversial, with a direct experimental measurement being absent. In this study, we utilise an advancement in non-linear vibrational spectroscopy to gain access to this important parameter. Combining phase-resolved sum- and difference-frequency generation (SFG and DFG) responses, we directly measure the decay in structural anisotropy of the air-water interface. We find a decay length of ~6-8Å, in excellent agreement with depth-resolved SFG spectra calculated from *ab initio* parameterised molecular dynamics (MD) simulations. The result reveals surprisingly short anisotropic orientational correlations from the interfacial layer that are even shorter than in the bulk. Furthermore, the recorded SFG and DFG responses are decomposed into a vibrationally resonant and non-resonant contribution through isotopic exchange measurements. Through their separate analysis, we show that the resonant response is a sensitive probe of the structural anisotropy at the interface whereas the non-resonant contribution contains a significant isotropic contribution from the bulk and therefore only partially reports on the interfacial structure. This finding places stringent restrictions on the insight available through both purely non-resonant and second-order intensity studies.

## Significance Statement

Aqueous interfaces are ubiquitous in many natural and artificial processes and their significance arises from the unique properties of water molecules within the interfacial region, with a crucial parameter being the thickness of its structural anisotropy, or "healing depth". Yet, experimental measurement of this quantity for the air-water interface remains elusive, with the current understanding mainly originating from molecular dynamics simulations. Using a novel vibrational spectroscopic tool, we provide the first measurement of this thickness and find it to be ~6-8Å. In combination with depth-dependent second-order spectra calculated from *ab initio* parameterised molecular dynamics simulations, which are in excellent agreement with this experimental result, we shed light on this surprisingly short correlation length of molecular orientations at the interface.

## Introduction

The air-water interface is ubiquitous in nature and serves as a useful model to study hydrophobic aqueous interfaces. Its importance is closely related to the unique characteristics of water in the interfacial region that is at the heart of numerous chemical processes in nature as well as industrial applications. This outstanding role of interfacial water has triggered an enormous number of experimental and theoretical investigations over the past several decades yielding exceptional insight into its structural properties.(1–7) Nevertheless, some of the most fundamental aspects of the air-water interface still remain controversial, particularly the length-scale or "thickness" of the interfacial region.(8)

The presence of the phase boundary makes the interfacial region anisotropic with physicochemical properties that strongly deviate from the corresponding bulk values. This anisotropy consists of depth-dependent variations in molecular density, dielectric constant, as well as the distribution of molecular orientations and the number, strength, and dynamics of hydrogen bonds within the intermolecular network. The details of these variations and the length-scale of their decay govern the specific role water plays in the function and behaviour of aqueous interfaces. For example, its dielectric properties influence its interactions with charges which play a role in chemical activity, ion transport, and electron transfer processes, while the density influences its viscosity and therefore many kinetic processes.(9–11) Some of the most pertinent properties of water, however, are controlled by the specific orientational distributions of the water molecules and details of the hydrogen bond network i.e., its molecular and intermolecular structure. These include its solvation behaviour and surface tension, which are critical in the thermodynamics underlying processes including uptake and transport mechanisms as well as chemical reactions.(12, 13)

Whilst the depth-dependent deviations in these different properties are all obviously interconnected, they can in principle decay on different length-scales. In consequence, any value of the anisotropic thickness that is experimentally measured depends on the specific property being probed. For the air-water interface, the anisotropy decay in density and dielectric constant have been experimentally determined using techniques such as neutron reflectometry(14) and ellipsometry(15, 16), respectively. These studies indicate that their variation occurs over length-scales of ~3-5Å, and thus that the bulk density and dielectric constant are recovered very quickly. In contrast, a direct experimental measurement of the thickness of the anisotropic structure (molecular orientations and intermolecular connectivity) remains elusive.(8)

The well-defined directionality and strength of hydrogen bonds in liquid water, along with its large molecular dipole, make the orientations of neighbouring water molecules highly correlated. In pure bulk water the length-scale of these correlations is, however, somewhat contentious owing to the many influencing factors. On the one hand, these correlations are often considered to be contained within length-scales of ~15Å, thus with angular reorientation events of individual molecules triggering the surrounding molecules within ~4-5 coordination shells to restructure.(17–20) On the other hand, they have also been indicated to extend much further to 10s or even 100s of Å through acoustic coupling and long-range dipole-dipole and orientationally restrictive hydrogen bonding correlations.(20–23) Such behaviour is also highly debated upon the addition of charged electrolytes which both perturb the dipole orientations and distort the hydrogen bond network.(22, 24–28) In any case, with correlations being present over several coordination shells and potentially much further, it is clear that the hydrogen bond network can have a vast reach in generating long-range molecular order. It is nevertheless unclear whether the specific anisotropic molecular structure present at the interface causes orientational correlations similar to those in the bulk, or if they are much longer, or even shorter.

Current insight into this question solely comes from molecular dynamics (MD) simulations which suggest a surprisingly short anisotropic structural thickness of ~6Å.(29–35) This would mean that the bulk structure is already obtained by the $4^{th}$, or even $3^{rd}$, molecular layer, and thus the reach of the hydrogen bond network of interfacial water on the molecular orientations below the interface is definitely shorter than in the bulk. This is especially interesting given that simulations have also indicated that the orientations of interfacial water molecules are actually highly correlated through expansive hydrogen bond connectivity, only that these correlations predominantly exist in-plane i.e., laterally within an overall isotropic 2D hydrogen bond network, and not normal to the interface.(36, 37) While this detailed structural view of the interface is very enlightening, it is also quite remarkable as it suggests that the interfacial molecules are somewhat detached from the bulk in an ultra-thin layer. However, as it is known that results and interpretations from MD simulations can be highly sensitive to their choice of forcefield and specific parameter-sets, it is crucial to confront these simulations with independent experimental verification of this rather unexpected result for the anisotropic thickness.

An experimental technique that has been widely applied to aqueous interfaces is vibrational sum-frequency generation (SFG) spectroscopy.(13, 38–53) The strength of SFG for experimentally addressing this question is its sensitivity to molecular orientations that are encoded in the sign of the output signal phase, as well as the molecular environments and intermolecular interactions that control the specific line-shapes of the vibrational resonances.(54) These can make it a selective probe of structural anisotropy as isotropic regions yield no signals under the electric dipole (ED) approximation owing to cancellation.(55, 56) SFG spectroscopy has been very successful in identifying specific interfacial water species such as water molecules with dangling bonds pointing into the air-phase ('free' OH).(38) However, extracting information about the thickness of the structural anisotropy with SFG spectroscopy has proven to be a veritable challenge. Recently, Benderskii et al. used isotopic dilution SFG experiments to investigate the intramolecular coupling between the dangling OH and its associated hydrogen bonded mode to assess the hydrogen bond strength of the latter, ultimately showing it to be almost equivalent to that in the bulk.(8) From this, they inferred that the structural anisotropy decays remarkably fast with depth. Later, Nagata et al. combined experimental measurements with simulations to investigate the anisotropy in the dielectric constant across the interface.(10) Through simulations of the different contributions to the structural anisotropy, they show that the free OH stretch contribution and those from the hydrogen bonded modes must arise from locations within the interface with differing dielectric properties. From their results, they determine that the variation in dielectric constant across the interface predominately occurs over ~1-3Å. However, the length-scale of the structural anisotropy in this study was entirely derived from simulation and not extracted from the experimental results. Nevertheless, the observation of differing dielectric environments for the different structural motifs does suggest a short decay length of the structural anisotropy.

There are two major obstacles for accessing the thickness of the structural anisotropy using SFG spectroscopy: i) the signals are integrated over depth and thus a single SFG measurement cannot directly yield information on this thickness, and ii) when probing water, it is still unclear to what extent the ED approximation holds and the measured signals really originate exclusively from structurally anisotropic regions. Beyond the ED approximation, signals can also be generated in the isotropic bulk (quadrupolar origin), which could easily represent a relevant contribution to the overall response. Whether such isotropic contributions are significant in the water response or not, has been a long-standing question in nonlinear spectroscopy and a clear answer has yet to be given.(57–60) Addressing this question is obviously essential for obtaining precise information on the structural anisotropy decay.

In this contribution, we utilise our recently developed depth-resolved vibrational spectroscopy that overcomes these limitations.(61, 62) This technique allows us to perform depth-resolved analysis of the structural anisotropy of the air-water interface and provide a direct measurement of its thickness. This is made possible through the simultaneous phase-resolved measurement of two different second-order responses, namely sum- and difference-frequency generation (SFG and DFG), which allows for both the precise depth profiling of the signal sources on the sub-nm scale, as well as an unambiguous quantification of isotropic signal contributions. The experimental results are then compared to depth-resolved SFG spectra from *ab initio* parameterised MD simulations. Furthermore, we show that, through isotopic exchange measurements, the overall nonlinear response can be decomposed into a resonant and non-resonant contribution. From their independent analyses, we then unravel their different spatial origins and discuss the far-reaching impact of this finding on second-order spectroscopy measurements.

## Results

In order to reveal the interfacial water structure and its evolution with depth, we use phase-resolved SFG-DFG spectroscopy across the ~2300-2900 cm$^{-1}$ frequency range to probe the O-D stretch vibration in $D_2O$. The general theory behind SFG spectroscopy can be found elsewhere in the literature(54–56, 63, 64), with the specifics underlying the depth-resolved SFG/DFG spectroscopy employed in this work detailed in previous publications(61, 62). Here we only provide a brief discussion of its main features.

The generation of the SFG and DFG (s-polarised) signals is achieved by nonlinear frequency mixing between two laser pulses, namely a mid-infrared (p-polarised) pulse that probes vibrational resonances and a visible upconversion (s-polarised) pulse with photon energy far from any sample resonance. The phases of these sample responses are determined interferometrically using SFG and DFG reference pulses (local oscillators, LO) that are linearly reflected by the sample surface. This plane of linear reflection (PLR) serves as reference position ($z = 0$) for our depth-resolved studies. Any second-order electric dipolar signal that arises from structural anisotropy and originates exactly from this depth plane, generates SFG and DFG spectra that precisely coincide in both phase and amplitude. However, signals from any deeper layers contributing to the dipolar response ($z > 0$) are phase-shifted in opposite directions for SFG and DFG (see Supplementary Information or references(61, 62) for details), as shown in Figure 1(a) for selected pathways. This phase-shift arises from the added propagation of the input and output beams and increases linearly with depth. The modulated phases lead to a phase difference between the integrated SFG and DFG responses that approaches 180º for large decay lengths, z', as shown in Figure 1(b). The apparent phase difference between SFG and DFG is hence a direct measure of the decay length of the structural anisotropy.

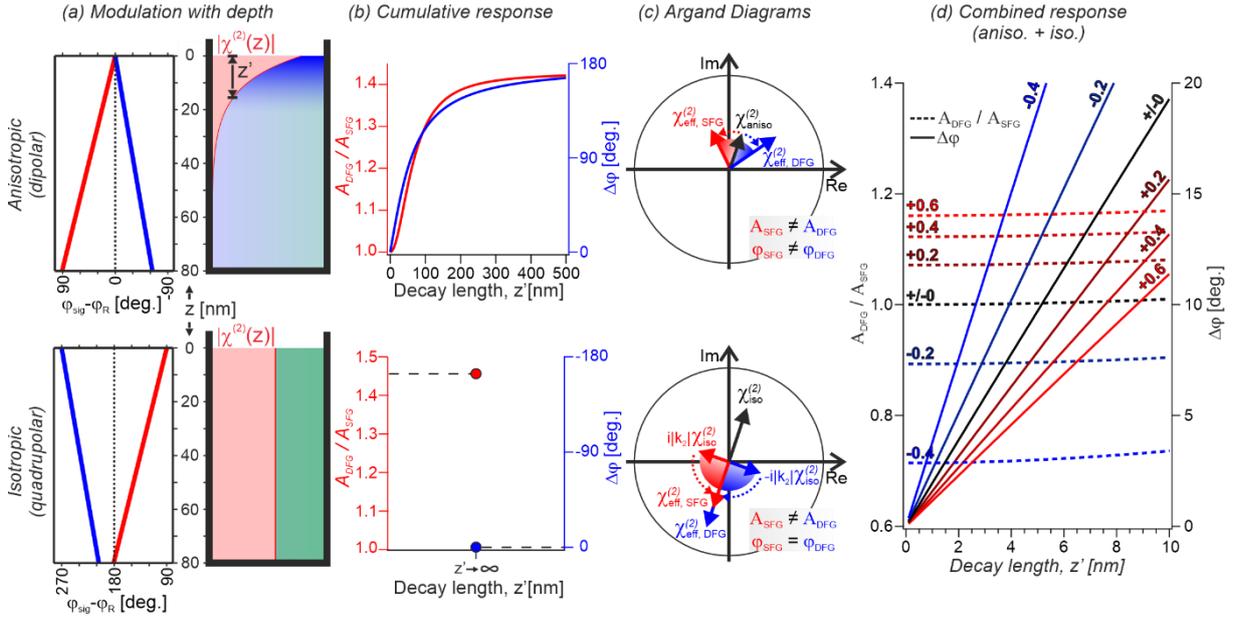

*Figure 1: Schematic representation of the amplitude and phase modulations to the SFG and DFG responses for both anisotropic (dipolar) and isotropic (quadrupolar) signals. (a) Graphical representations of the modulation of the phase and amplitude of the responses from chromophores as a function of depth. $\phi_{sig}$ represents the phase of the output signal from each depth and $\phi_R$ represents the resonant phase. (b) Graphs of the phase difference, $\Delta\phi = \phi_{SFG} - \phi_{DFG}$, and amplitude ratio, $A_{DFG}/A_{SFG}$, of the cumulative (depth-integrated) responses as a function of the decay length of the signal contribution, z'. As the isotropic contribution does not decay (part (a), bottom-right panel), $z' \to \infty$ and only the limiting values are presented. (c) Schematic Argand diagrams of the effect of depth on the phase and amplitude of the effective (measured) SFG and DFG responses, $\chi_{eff}^{(2)}$. (d) Plot of the amplitude ratio and phase difference of the combined response comprising an anisotropic contribution and varying relative isotropic contributions. The ratio of the amplitudes of the isotropic to anisotropic contributions are indicated on each trace.*

In contrast to the anisotropic (dipolar) response, any contribution arising from isotropic environments (quadrupolar sources) has distinctly different characteristics. As shown in the Supplementary Information, this intrinsic response is inherently shifted by +90º for SFG and -90º for DFG. The corresponding integration over depth further phase-shifts the SFG and DFG responses, leading to a decreasing phase difference that tends to 0°. As the isotropic signals always originate from the entire bulk, the phase difference between their integrated SFG and DFG responses must be precisely zero, unlike the dipolar case. In addition to the phase-shift, any integration also leads to differing amplitudes for SFG and DFG, which only become significant for depth values >>10 nm (see Supplementary Information). Such a situation is obviously given for any isotropic contribution (as $z' \to \infty$) but typically not the case for the anisotropic response when considering its expected nanoscale decay length. Generally, the anisotropic contribution can, in principle, yield different phases and amplitudes for SFG and DFG, however, only the phase difference is typically significant. On the other hand, the isotropic contribution presents no phase difference but an amplitude ratio clearly deviating from unity.

Based on these differing characteristics, anisotropic and isotropic signal sources can be separated, and the purely anisotropic decay can be determined. Particularly straight forward is this determination for the typical cases where the anisotropy decay is rather small (<10 nm) as depicted in the graphical representation in Figure 1(d). Here, the theoretical overall amplitude ratio and phase difference are shown as a function of the anisotropic decay length considering different relative isotropic contributions to the combined response. From this, it becomes clear that the deviation of the amplitude ratio from unity exclusively speaks to the relative proportion of isotropic component, while the phase difference is modulated by both aniso- and isotropic responses. Therefore, the exact decay length of the structural anisotropy can be extracted by first determining the isotropic contribution based on the amplitude ratio and using this to correctly transform the measured phase difference into the corresponding value of z'.

With this analytical procedure in hand, we turn to the experimental results from the $D_2O$-air interface shown in Figure 2(a). The obtained SFG and DFG spectra are split into their real and imaginary parts, corresponding to the dispersive and absorptive line-shapes, respectively. Before performing the depth analysis, we briefly discuss the obtained resonant line-shape. The spectra exhibit four clearly distinguishable absorption bands, two negative contributions at ~2400 and 2540 $cm^{-1}$ which highly overlap, along with two positive bands, one being a sharp feature at 2740 $cm^{-1}$, and a broader shoulder to this band at ~2680 $cm^{-1}$. The specific resonant frequencies of the stretching modes are particularly sensitive to hydrogen bond strength, with them becoming increasingly red-shifted for stronger intermolecular bonds.(65) Therefore, each structural motif in liquid water possesses a characteristic vibrational response which enables their identification. The positive sharp band at 2740 $cm^{-1}$ is assigned to the free OD stretch where the positive sign of the resonant peak indicates a preferential "pointing up" orientation of this water species, in accordance with its interpretation.(65–68) By contrast, the two overlapping negative bands between 2300 and 2600 $cm^{-1}$ originate from hydrogen bonded O-D stretch vibrations having transition dipoles pointing down on average.(65–68) Furthermore, the assignment of the positive shoulder at 2680 $cm^{-1}$ overlapping with the free OD has been highly contentious over the past several years, although the positive sign indicates it has a general direction towards the air-phase. Its origin has been suggested as the antisymmetric OD stretch arising from intramolecular coupling from $D_2O$ molecules presenting one acceptor and two donor hydrogen bonds(8, 46, 69), but more recently has been assigned to a Fermi resonance of the free OD with a combination band mixing the hydrogen-bonded OD stretch with a low-frequency intermolecular vibration(70). Finally, in addition to the resonant line-shape, it is important to note the presence of a significant non-resonant contribution arising from electronic interactions which can clearly be seen by the large negative offsets in the real parts of the spectra.

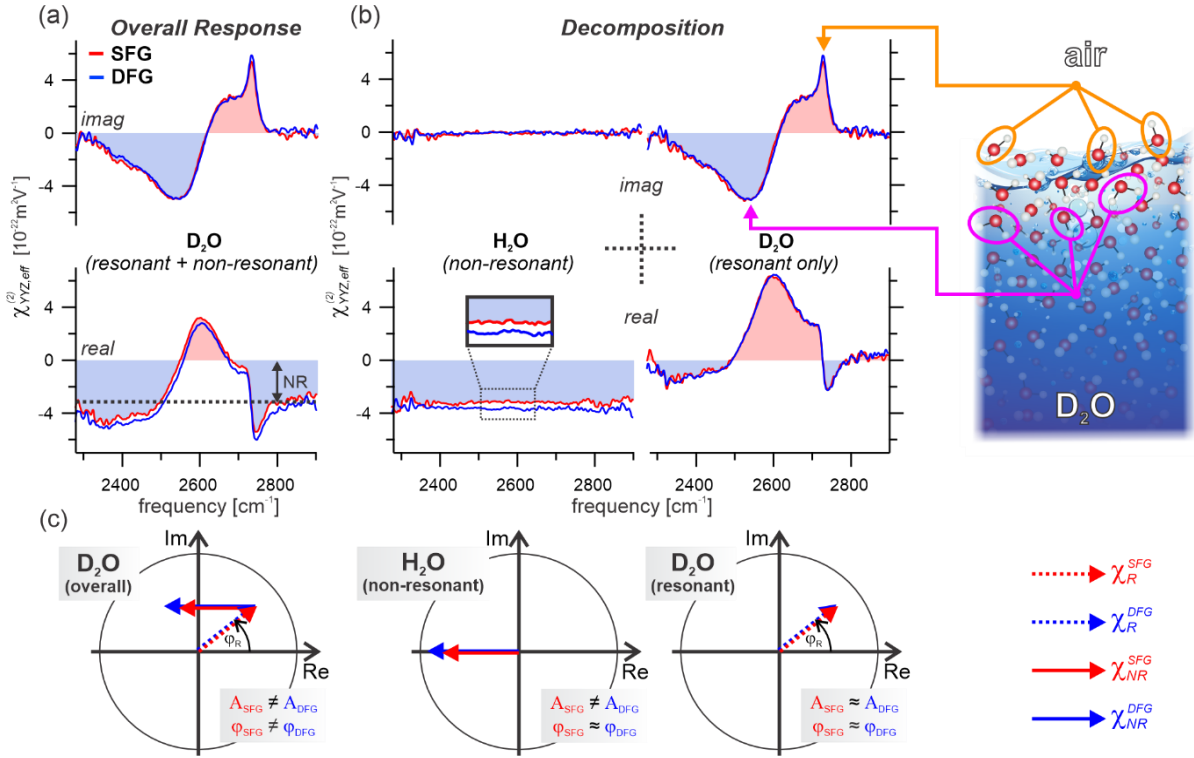

*Figure 2: SFG and DFG spectra of the air-water interface in the SSP polarisation combination. (a) real and imaginary parts of the second-order response from $D_2O$ given in absolute units. The dashed line in the real part indicates the non-resonant contribution to the spectra (NR). (b) real and imaginary parts of the purely non-resonant $H_2O$ response as well as the $D_2O$ response having subtracted that from $H_2O$, thus only representing the resonant contribution. The spatial origin of structural motifs giving rise to the two most significant stretching bands, namely the "free" OD at 2740 $cm^{-1}$ and H-bonded OD at ~2540 $cm^{-1}$, are schematically indicated. (c) Schematic Argand diagrams of the three spectra presented in (a) and (b), emphasising any differences between the amplitudes and phases of the SFG and DFG responses.*

Upon comparison of the SFG and DFG spectra, they show remarkable similarities, but are clearly not identical. This immediately suggests the presence of signal contributions from non-zero depth. On closer inspection, the vast majority of the difference is present in the real parts (Figure 2(a), lower panel) which appear to only differ by a constant offset. Interestingly, this suggests that the difference arises solely from the non-resonant contribution. Apparently, the resonant and non-resonant contributions seem to report on different depth profiles, necessitating their separation. This is achieved through isotopic exchange experiments by measuring the analogous spectra for $H_2O$. Since both isotopologues have identical electronic structures, it is reasonable to treat their non-resonant contributions, which are dominated by electronic interactions, as being equal. The $H_2O$ spectra are depicted in Figure 2(b), showing that they indeed precisely reproduce the same apparent negative offset as in $D_2O$. Based on this, the overall spectra can be fully decomposed into their pure resonant and non-resonant contributions, as shown in Figure 2(b).

With the resonant and non-resonant contributions separated, we see that the purely resonant SFG and DFG spectra (Figure 2(b), right-side) almost perfectly overlap, and thus that this contribution reports on a short anisotropic decay (small phase difference) and has no significant isotropic contribution (equal amplitudes). The purely non-resonant spectra also show little phase difference (Figure 2(b), left-side, imaginary parts are both ~0), but, in contrast, clearly feature a deviation in their amplitudes (Figure 2(b), left-side, offset in real part). Following Figure 1(d), this demonstrates that the non-resonant contribution must possess a considerable isotropic component. For its quantification within both signal contributions, we assess the amplitude ratios from each. For the non-resonant contribution, this can be done with high precision as it is independent of frequency and thus can be spectrally fitted with a constant value. The obtained ratio of 1.1480±0.0022 reveals that ~34% of the SFG (and ~42% of the DFG) non-resonant response originates from the isotropic bulk, representing a remarkably large bulk contribution (see Supplementary Information for details). In contrast, the average value of the amplitude ratio for the resonant contribution is 1.00±0.04. This mean value of precisely 1 indicates that there is no considerable isotropic contribution, consistent with the observation of highly overlapping spectra. However, the larger standard deviation compared to the non-resonant contribution reports a larger uncertainty for this assessment. Nevertheless, the size of the standard deviation in the average value can be used to put an upper bound on a possible isotropic contribution, showing that it must be <10% of the overall resonant response. Therefore, the resonant contribution is clearly dominated by the anisotropic dipolar signal. The observed differences between the SFG and DFG resonant and non-resonant contributions are well-represented by the schematic Argand diagrams in Figure 2(c).

Evidently the determination of the isotropic contributions, and thus the anisotropic depths, is highly sensitive to the accuracy at which the amplitude ratio is determined. Therefore, any possible other sources of deviations, such as dispersion effects must be ruled out. The energy level diagram in Figure 3(a) shows that SFG and DFG involve different frequencies, and thus dispersion could potentially by present, although is typically insignificant.(61) Nevertheless, we test experimentally for any impact from dispersion by measuring a separate DFG response (labelled DFG', see Figure 3(a)) using a shifted upconversion at the original SFG frequency to compare to the original DFG response, as shown in Figure 3(b). If dispersion is zero, the amplitude ratio between the two non-resonant responses should be close to unity, with a slight deviation arising from a small modulation of the apparent isotropic contribution (see Supplementary Information). The theoretically derived ratio for this case is 0.988. The presence of dispersion effects should meanwhile appear as a clear deviation from this value. From a comparison of the fitted values from the experimental results, we obtain a measured amplitude ratio of 0.991, which is in remarkable agreement with the predicted value. A summary of this comparison is given in Table 1. These values clearly show that dispersion effects are indeed negligible in these experiments.

*Table 1: Comparison between measured and predicted values of the ratio of $H_2O$ non-resonant amplitudes for DFG responses measured with upconversion beams at 585 and 690 nm. The uncertainty in the measured value arises from the fit of the non-resonant response whilst that for the predicted value sources from the uncertainty in the relative isotropic contribution, and thus from the measured amplitude ratio between SFG and DFG.*

| Susceptibility Ratio | Measured | Predicted |
|---|---|---|
| $\left\|\dfrac{\chi_{eff}(DFG')}{\chi_{eff}(DFG)}\right\|$ | 0.991±0.003 | 0.988±0.010 |

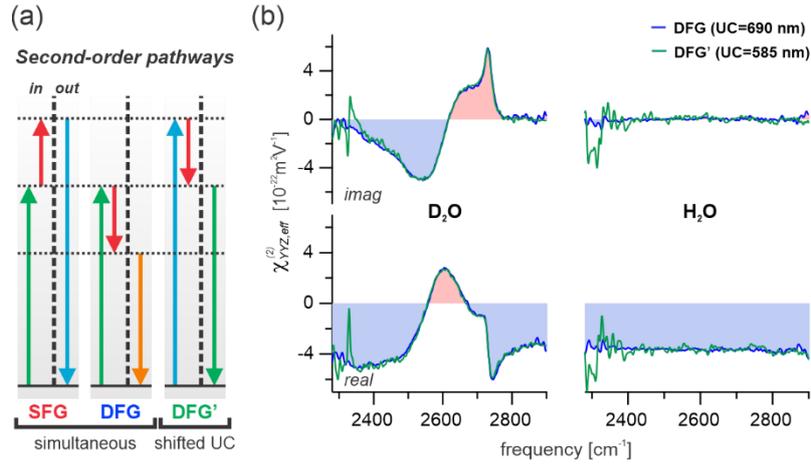

*Figure 3: Dispersion test for the observed amplitude ratio in the non-resonant response. (a) Schematic energy level diagrams of the SFG and DFG pathways produced simultaneously as well as for a separate DFG pathway, DFG', using a shifted upconversion frequency at the initial SFG frequency. (b) Comparison of the resulting DFG and DFG' spectra for both $D_2O$ and $H_2O$.*

Based on these findings, we can now extract the decay length of the structural anisotropy from the phase difference between SFG and DFG depicted in Figure 4(a). The resonant phase difference spectrum is very close to zero, but generally slightly positive, which corresponds to locations just below the interface. The average values of the phase differences across all frequencies within the bandwidth of the resonances are 1.50±0.10° for the resonant contribution and 0.36±0.10° for the non-resonant contribution. With their respective proportions of isotropic contributions given above, these phase differences can be converted into their corresponding decay lengths of the anisotropic contributions, yielding values of 7.7±1.0Å for the resonant component and 3.1±0.9 Å for the non-resonant component. It is important to note that the stated uncertainties in these values are derived from the combination of the uncertainty in the fit for the phase difference and the uncertainty in the conversion factor between phase difference and decay length which stems from the measured size of the isotropic contribution. They are hence neglecting any systematic errors as well as any inherent frequency dependence to the phase difference, which could be present in the resonant contribution. Therefore, in reality, the confidence intervals of the decay lengths for each contribution are likely broader and could well span several Ångströms. Despite this uncertainty, with both values being clearly below 1 nm, it can be safely concluded that no significant structural anisotropy extends beyond the first ~3 molecular layers.

The experimentally obtained mean value for the decay length of the resonant response (7.7Å) can be directly compared to the predictions from MD simulations (which only include the vibrationally resonant contributions). The right panel of Figure 4(b) shows the depth-dependent second-order susceptibility in absolute units extracted from MD simulations. A quantitative comparison of the overall integrated response with the experimentally obtained purely resonant spectrum is then presented in Figure 4(c). The simulation results clearly show the same resonant features as in the experiment, namely the two overlapping negative (blue) low-frequency modes as well as the positive (red) free OD stretch at 2940 cm$^{-1}$ and its low-frequency shoulder, along with excellent agreement in absolute amplitudes. Figure 4(b) then reveals that both positive and negative spectral features source from essentially the same interfacial region, with equal onsets and termination depths. On closer inspection within this region, there is a clear red-shift with increasing depth for both spectral features, reflecting the gradient in hydrogen bond strength. Overall, the simulated resonant features are contained within a thickness of 6Å (from $z = -2$Å down to 4Å). A direct comparison to the experimentally obtained value (~7.7Å) places the experimental and simulation results within 2Å of each other i.e., within a single molecular layer, and thus in consensus within the uncertainty of each method.

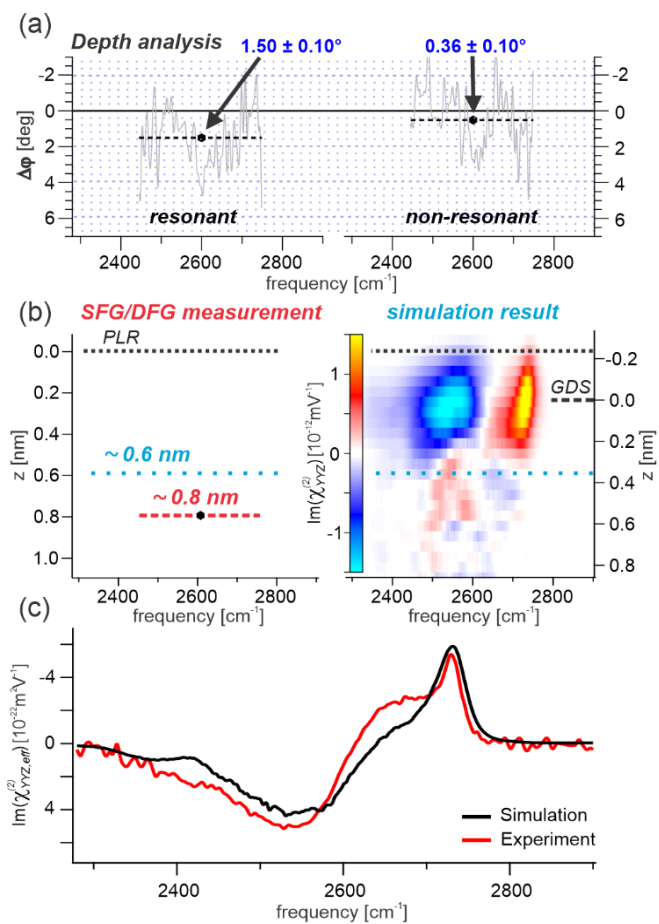

*Figure 4: Depth analysis of the resonant and non-resonant responses. (a) Plotted phase difference between SFG and DFG for each contribution. The raw data in each plot have been fitted with a constant to extract a specific depth value. (b) Comparison between the experimentally obtained depth value and that from depth-dependent SFG spectra calculated from ab initio parameterised MD simulations. (c) Integrated SFG response overlapping with the experimentally obtained purely resonant SFG $D_2O$ spectrum, with both shown in absolute units. (PLR: plane of linear reflection, GDS: Gibbs dividing surface. Note the different zero positions, see Supplementary Information for details.)*

## Discussion

As shown above, the resonant and non-resonant contributions report slightly different mean values for the anisotropy decay (7.7Å versus 3.1Å). Whilst this discrepancy likely falls within their respective confidence intervals, there is a notable difference. This raises the question of which of the values more accurately describes the structural anisotropy. The resonant response is dominated by vibrational interactions and hence should be highly sensitive to both the distribution of molecular orientations and any anisotropy in the hydrogen bond connectivity. The sensitivity to the former arises from the substantial asymmetry in the Morse potential, whilst the latter is due to the strong frequency dependence of the OD stretch vibrations on their hydrogen bond environment.(71) Furthermore, the resonant response contains no significant isotropic quadrupolar contributions. Clearly, therefore, any anisotropy of the molecular structure within the extensive hydrogen bond network is expected to be the predominant factor influencing the length-scale obtained from the resonant response. On the other hand, the non-resonant response is dominated by electronic interactions and is thus generally related to the asymmetry of the electron cloud(72) and could well be less sensitive to molecular orientations and any intermolecular interactions, and thus show a different evolution with depth. Here, it is shown that the water non-resonant response even contains substantial isotropic contributions which compose almost half of the overall signal. Thus, a significant portion of the signal is clearly not reporting on any anisotropic aspects of the interface. Furthermore, it is even possible that the overall non-resonant response is highly insensitive to structural anisotropy altogether. Given its substantial isotropic contribution from bulk quadrupolar sources, is it not unreasonable to expect that anisotropic (interfacial) quadrupolar sources may also be significant contributors since they originate from the same fundamental mechanism.(58) Unlike the dipolar mechanism which solely reports on the structural anisotropy, the anisotropic quadrupolar contributions arise from the discontinuity of fields at the interface and thus primarily report on the length-scale of the dielectric anisotropy.(55) In this context, the value of ~3.1Å obtained here agrees remarkably well with previous measurements of the thickness of the dielectric variation across the interface.(9) It is thus entirely possible, that the dipolar contribution only represents a minor contributor to the overall non-resonant response of water.

Based on the discussion above, we conclude that the resonant response is indeed a far better probe of the structural anisotropy. Furthermore, as the extracted length-scale is in good agreement with the simulation result, both experiment and theory report a complementary view of the structural evolution of water at the interface with air. Hence, it can be conclusively stated that the effect of the phase boundary on the out-of-plane molecular structure does diminish remarkably quickly (~6-8 Å) and seems to concern only the first 3 molecular layers. This means that the scale of the anisotropic structure (both orientational correlations and hydrogen bond connectivity normal to the interface) of interfacial water is rather shorter than the length scale of the correlations in the isotropic bulk. Whilst this result follows theoretical predictions well(29–37), it is still somewhat surprising and raises questions about the underlying factors that dominate the structural anisotropy at the interface. Firstly, the hydrogen bond connectivity is clearly reduced / weakened in the first molecular layer at the interface, making it unfavourable in terms of its free energy (as evidenced by the considerable surface tension of water) and thus there is a significant driving force towards retaining the bulk connectivity as quickly as possible. From this perspective, the observed fast decay is in line with expectations. This driving force, however, does not necessarily impose a loss in any preferential orientation. Of course, the lower connectivity of the interfacial molecules suggests that they inflict less orientational restriction on subsequent layers, but such an effect only considers the impact of individual molecules, and not their cumulative alignment and the resulting electrostatic effects from oriented dipoles. If the interface induces a substantial out-of-plane preferential dipolar direction, even if this oscillates between consecutive molecular layers, the alignment of dipoles could be expected to impose similar preferential orientations in the layers beneath, and thus exhibit longer range correlations. On the other hand, the lower hydrogen bonding connectivity at the interface enables greater orientational freedom, and thus a gain in entropy, in agreement with the well-known effect of decreasing surface tension on increasing temperature. Such an entropic gain can, however, only be realised through short orientational correlations. Therefore, since larger correlations are not observed, this suggests that the entropic gain dominates over the electrostatic correlation effects, and thus that the hydrogen bond network is the predominant factor influencing the structural anisotropy at the interface.

Beyond these insights into the water structure, our findings also have far-reaching consequences for non-linear optical measurements on aqueous interfaces. As shown above, the non-resonant response of water is both clearly not a selective probe for anisotropic environments and may well be even fairly insensitive to structural anisotropy. Since a central pillar of second-order measurements is the anisotropic selectivity, these findings place substantial constraints on the interpretation of second-order spectra from aqueous interfaces. This concerns, in particular, intensity SFG approaches where the resonant contribution cannot be isolated, and second harmonic generation (SHG) spectroscopy that entirely relies on the interpretation of non-resonant signals. In contrast, the resonant response has been demonstrated to be a good marker for structural anisotropy, and has been shown to be extremely localised to the surface. This makes both amplitude and phase of the resonant response highly insensitive to the specific experimental setting, with any effects from incident beam angle or Fresnel factors being separable from the measured response. The reported resonant spectrum can thus be considered as an intrinsic property of the water surface and could be a useful reference for comparisons between different experimental set-ups as well as for the result of simulations. On the other hand, the non-resonant response will change with experimental settings due the combination of isotropic and anisotropic contributions (as discussed in Supplementary Information). It is therefore not an intrinsic property of the system. However, this dependency is only manifested in its amplitude, with its phase being highly insensitive to the specific settings. This is due to its isotropic contribution being entirely real and the anisotropic component reporting a very small depth. This result is especially important as the potential role of the non-resonant water response as a phase reference in second-order measurements has been controversial, with no agreement on its true phase.(73–75) Nevertheless, our results conclusively show that the phase of the non-resonant response is very close to $\pm 180°$ (specifically, -179.6° for SFG) and is almost entirely insensitive to the specific experimental settings. This contrasts strongly to quartz which is by far the most commonly used phase and amplitude reference, but has been shown to have a significant phase deviation from the typically assumed phase of $\pm 90°$.(76, 77) On the other hand, unlike quartz, the non-resonant water response is clearly not a good amplitude reference. As such, the combination of water and quartz would make an excellent reference pairing for phase and amplitude measurements, respectively.

## Conclusions

In conclusion, we report the experimentally determined thickness of the structural anisotropy of the air-water interface using a newly developed second-order vibrational spectroscopy, finding it to be ~6-8Å. The obtained decay length is compared to depth-resolved SFG spectra calculated from *ab initio* parameterised MD simulations, showing excellent agreement. These combined results report on a remarkable short length-scale for both the correlation of molecular orientations normal to the interface and recovery of bulk hydrogen bond connectivity, covering only 3 molecular layers. This ultra-short correlation length-scale highlights the important role of the interfacial entropy alongside the loss in hydrogen bonding connectivity in dictating the molecular structure at the interface. Furthermore, we show that the resonant signal from the OD stretch vibration is indeed a selective probe of the structural anisotropy whereas the non-resonant (electronic) contribution is found to be little selective as it contains significant contributions from the isotropic bulk. This result underlines the importance of a careful analysis of the mechanistic origin of the signals in second-order spectroscopy and raises fundamental questions about the correct interpretation of results from non-resonant studies of aqueous interfaces. This also includes resonant SFG studies if the non-resonant contribution is not properly accounted for. Nevertheless, we have demonstrated that the presented depth-resolved vibrational spectroscopic technique allows these challenges to be overcome and obtain precise insight into the evolution of the structural anisotropy with depth in such systems.

## Acknowledgements

The authors would like to acknowledge the Deutsche Forschungsgemeinschaft (DFG) for funding (Project-ID 221545957 - SFB 1078/C1).

## Competing Interests
There are no competing interests.

## Author Contributions
M.T. conceived the project and A.P.F., Á.D.D., V.B. and M.T. designed and performed the experiments while L.L. and R.R.N. undertook the simulations. Analysis of the experimental data was performed by A.P.F. and M.T. and all authors discussed the interpretation of the results. A.P.F. drafted the manuscript which was edited by all authors. R.R.N., M.W., and M.T. supervised the work and acquired funding.

## Data Availability
Raw data will be made available upon reasonable request by contacting the corresponding author.

nonlinear bulk and surface response from alpha-quartz using phase sensitive SFG spectroscopy. *J. Chem. Phys.* **151**, 64707 (2019).

# How Thick is the Air-Water Interface? - A Direct Experimental Measurement of the Decay Length of the Interfacial Structural Anisotropy

## Supplementary Information

Alexander P. Fellows[1], Álvaro Díaz Duque[1], Vasileios Balos[2], Louis Lehmann[3], Roland R. Netz[3], Martin Wolf[1], and Martin Thämer[1]*

[1] Fritz-Haber-Institut der Max-Planck-Gesellschaft, Faradayweg 4-6, 14195, Berlin, Germany

[2] Instituto Madrileño de Estudios Avanzados en Nanociencia (IMDEA Nanociencia), 28049, Madrid, Spain

[3] Department of Physics, Freie Universität Berlin, Arnimallee 14, 14195, Berlin, Germany

\* Corresponding author

thaemer@fhi-berlin.mpg.de

(tel.): +49 (0)30 8413 5220

# Materials and Methods

## Sample Preparation

The spectroscopic measurements of the air-water interface were performed on both $H_2O$ (Milli-Q, 18.2 MΩ·cm, <3ppb TOC) and $D_2O$ (VWR Chemicals, 99.9% D). The water was contained in a custom-made Teflon (PTFE) trough which was cleaned overnight with Piranha solution (3:1 sulfuric acid to 30% hydrogen peroxide solution) and thoroughly rinsed with ultrapure water prior to use. *Warning: Piranha solution is highly corrosive and an extremely powerful oxidiser. Great care must be taken with its preparation and use.*

## Spectral Acquisition

The SFG and DFG spectra were recorded in the time-domain using a home-built non-linear interferometer, the details of which can be found elsewhere.(1) In short, two 1 kHz 800 nm outputs (4 and 3W) of a Ti:Sapphire laser (Astrella, Coherent) are used to pump two optical parametric amplifiers (TOPAS, Light Conversion). In the first, the pump beam is split into signal and idler, with the signal output being taken and frequency-doubled using a BBO crystal to use as the upconversion beam. The second TOPAS is used to generate tuneable mid-IR via a DFG unit. Part of the IR beam (~5%) is taken off using a beamsplitter and combined collinearly with the upconversion, to generate local oscillator (LO) beams from z-cut quartz. The fully collinear output containing the upconversion and both SFG and DFG reference beams (LOs) are then recombined collinearly with the remaining ~95% of the IR after an interferometric translation stage to control the relative timing of the pulses. The combined beam is then sent towards the sample via an oscillating mirror to split the beam in two and perform shot-to-shot referencing using a reference z-cut quartz crystal. Both beam paths are focussed towards the samples at a 70° incidence angle from the surface normal after which they are recombined, spectrally filtered and detected using silicon photodiodes implementing balanced detection.

The spectra were measured with fast-scanning over the time delays of -500 to 6000 fs to ensure sufficient frequency resolution in the resulting spectra. The presented spectra using the 690 nm upconversion represent the average across 30,000 measurements taken from 3 different samples and those using the 585 nm upconversion represent the average from 20,000 measurements across two different samples. The spectra from each sample were compared and showed excellent reproducibility. Spectra were recorded in the ~2300-2900 $cm^{-1}$ frequency range, thus covering the O-D stretching region. This region was selected instead of the O-H stretching region purely for experimental reasons. Firstly, the IR generation from the TOPAS is significantly more efficient in the O-D region, thus giving access to greater IR powers. Secondly, the suppression of any parasitic contributions from the collinear beam geometry requires less optical material in the lower frequency range, and is thus easier to implement and ensure good quality spectra. Finally, due to the effective mass difference, the O-D stretches cover a narrower frequency range than the O-H stretches. Therefore, covering the entire region within the envelope of the IR is more straightforward and yields better signal-to-noise across the entire resonant line-shape, especially given that the IR bandwidth generated from the TOPAS is larger at lower frequencies.

During measurement, the entire optical path up to and including the sample is purged with dry, $CO_2$-free air to minimise any atmospheric absorption. Additionally, to ensure no change in the beam path or relative position of the sample surface, the height of the sample is continuously corrected for evaporation using an automated z-stage. Given that the measurement of the air-water interface necessitates a pure, clean surface, it is imperative that no surface contamination affects the results. However, even if some contamination to the surface occurs during the measurements, the local heating from the IR beam in the vicinity of the laser spot causes any surface-active species to migrate away via a substantial Bénard-Marangoni force.(2) Therefore, only at relatively high surface concentrations would any contaminants enter the probed surface region and thus alter the obtained spectra.

*Amplitude and Phase Correction*

The acquired SFG and DFG spectra were referenced in amplitude using the spectra from a z-cut quartz sample to remove the effect of the IR envelope. They were then further corrected for Fresnel factors and the beam geometry and converted into absolute units using the known susceptibility of quartz (0.6 pmV$^{-1}$)(3). This quartz measurement also gives an absolute phase reference for the sample spectra, which was taken to be ±90°, assuming the signal from quartz is an entirely non-resonant bulk dipolar response starting from the surface. As discussed in the main text, however, although this amplitude correction is valid, the assumed phase from quartz is slightly inaccurate.(4) For this reason the phases were corrected using a further SFG/DFG measurement of an octadecyltrichlorosilane (OTS) monolayer self-assembled on fused silica (FS). Given that the signals arise from the terminal methyl groups in such a sample, they effectively have no depth and thus the phase of their SFG and DFG response should be precisely equal. A more detailed discussion of this phase correction is given below when comparing zero positions for the depth.

*Calculation of Spatially Resolved SFG Spectra from Molecular Dynamics Simulations*

The theoretical prediction of SFG- and DFG spectra is based on the second-order polarization given in Eq. 1 (5–7)

$$\frac{1}{\varepsilon_0} P_i^{(2)}(z,t) = \int_{-\infty}^{\infty} \frac{d\omega_1}{2\pi} \int_{-\infty}^{\infty} \frac{d\omega_2}{2\pi} e^{-i(\omega_1+\omega_2)t} \chi_{ijk}^{(2)}(z,\omega_1,\omega_2) F_j^1(\omega_1) F_k^2(\omega_2) \qquad (1)$$

where $\chi_{ijk}^{(2)}(z,\omega_1,\omega_2)$ is the second-order response function and $F_j^1(\omega_1)$ and $F_k^2(\omega_2)$ are external electric fields which represent D or E fields (8, 9). We employ the Einstein sum convention and $i,j,k \in \{x,y,z\}$ are Cartesian indices. As the system is homogeneous in the xy-plane, z-polarized external fields correspond to electric displacement fields $\varepsilon_0^{-1} D_z(t)$ and x- or y-polarized fields correspond to electric fields $E_{x/y}(t)$. As we are interested in the nonlinear response of the system to mono-chromatic fields, we use Eq. 2

$$F_i^\beta(\omega) = 2\pi \mathcal{F}_i^\beta \delta(\omega_\beta - \omega) \qquad (2)$$

where $\mathcal{F}_i^\beta$ is the vectorial amplitude of the external field and the index $\beta \in \{IR, VIS\}$ distinguishes the IR from the visible (VIS) field source. As the VIS field does not resonate with the system, the dependence of the nonlinear response function on $\omega_{VIS}$ can be neglected, i.e., $\chi_{ijk}^{(2)}(z,\omega_{VIS},\omega_{IR}) \approx \chi_{ijk}^{(2)}(z,\omega_{IR})$. Consequently, we can write the nonlinear response of the system as in Eq. 3.

$$\frac{1}{\varepsilon_0} P_i^{(2)}(z,t) = e^{-i(\omega_{IR}+\omega_{VIS})t} \chi_{ijk}^{(2)}(z,\omega_{IR}) \mathcal{F}_j^{VIS} \mathcal{F}_k^{IR} \qquad (3)$$

In the SSP polarization combination, the corresponding position and frequency-dependent susceptibility for SFG and DFG is defined by Eq. 4

$$\chi^{(2)}_{yyz}(z, \omega_{IR}) = \varepsilon_{zz}^{-1}(z, \omega_{IR})\hat{\chi}^{(2)}_{yyz}(z, \omega_{IR}) \tag{4}$$

where $\varepsilon_{zz}^{-1}(z, \omega_{IR})$ is the inverse dielectric profile, which can be extracted by methods described earlier(10). The difference between the response function $\chi^{(2)}_{yyz}(z, \omega_{IR})$ and the susceptibility $\hat{\chi}^{(2)}_{yyz}(z, \omega_{IR})$ is that the former is a response function to general external fields, while the latter is a response function to electric fields. Ultimately, the experimental signal is determined by the integral over $\chi^{(2)}_{yyz}(z, \omega_{IR})$ and not $\hat{\chi}^{(2)}_{yyz}(z, \omega_{IR})$. However, remaining parts of this work are formulated with respect to the susceptibility $\hat{\chi}^{(2)}_{yyz}(z, \omega_{IR})$ and thus the difference needs to be clarified. Assuming classical nuclei motion, the imaginary part of the response function $\chi^{(2)''}_{ijk}(z, \omega_{IR})$ is given by the fluctuation-dissipation relation shown in Eq. 5.

$$\chi^{(2)''(z,\omega_{IR})}_{ijk} = \frac{\omega_{IR}}{2\tau_{max}\varepsilon_0 k_B T L_y L_x} a_{ij}(z, \omega_{IR}) M_k(\omega_{IR})^* \tag{5}$$

Here $a_{ij}(z, \omega_{IR})$ is the frequency-dependent polarizability profile, $M_z(\omega_{IR})^*$ is the complex conjugate of the frequency-dependent polarization of the entire system, $k_B$ is the Boltzmann constant, $T$ is the temperature, $\tau_{max}$ is the length of the trajectory, $L_x, L_y$ are the box dimensions in the plane of the interface. The trajectories are generated with the highly accurate MB-pol (11–13) force field with classical nuclei dynamics using LAMMPS(14). The polarization is computed with the Thole-type polarizability model TTM-4F (15), included in MB-pol. The molecular polarizabilities are parameterized from single-molecule *ab-initio* calculations on the CCSD(T)/aug-cc-pVTZ level, using Gaussian 16 (16). Here we expand the molecular polarizability tensor in the molecular Eckart frame $\boldsymbol{\alpha}_{mol}(S_1, S_2, S_3)$ to first order in the symmetry coordinates $S_1, S_2, S_3$ (17). Accordingly, the time-dependent polarizability of the $n^{th}$ molecule in the laboratory frame is given by Eq. 6

$$\boldsymbol{\alpha}(t) = \boldsymbol{R}[\boldsymbol{\Omega}_n(t)]^T \boldsymbol{\alpha}_{mol}[S_1^n(t), S_2^n(t), S_3^n(t)] \boldsymbol{R}[\boldsymbol{\Omega}_n(t)] \tag{6}$$

where $\boldsymbol{\Omega}_n(t)$ and $S_1^n(t), S_2^n(t), S_3^n(t)$ are the orientation of the Eckart frame and the symmetry coordinates of the $n^{th}$ molecule, respectively, and $\boldsymbol{R}[\boldsymbol{\Omega}_n(t)]$, is a rotation matrix. The influence of intermolecular interactions is accounted for by solving the self-consistent equation shown in Eq. 7 in each step iteratively.

$$\delta\boldsymbol{\mu}_n(t) = \boldsymbol{\alpha}(t)\left(E_n^P[\delta\boldsymbol{\mu}_N(t)] + \delta\boldsymbol{F}\right) \tag{7}$$

Here $\delta\boldsymbol{\mu}_n$ is the induced dipole moment of the $n^{th}$ molecule due to an external field $\delta\boldsymbol{F}$ and the field due to the induced dipoles on the other molecules $E_n^P[\delta\boldsymbol{\mu}_N(t)]$. The effective polarizability of the $n^{th}$ molecule is then given by Eq. 8

$$\alpha_{n,ij}^{eff}(t) = \frac{\delta\mu_{n,i}(t)}{\delta F_j} \tag{8}$$

where $\delta\mu_{n,i}(t)$ and $\delta F_j$ are the $i$ and $j$ components of the induced dipole moment and external field, respectively. Thus, the z-resolved polarizability profile of the system is given by Eq. 9

$$\mathbf{a}(z, \omega_{IR}) = \int_{-\tau_{max}/2}^{\tau_{max}/2} dt\, e^{i\omega_{IR}t} \sum_{n=1}^{N} \boldsymbol{\alpha}_n^{eff}(t)\delta[z - z_n(t)] \tag{9}$$

where $z_n(t)$ is the z-component of the centre of mass of the n$^{th}$ molecule. We bin all profiles with a bin size of 0.05 nm. In contrast to previous works (18, 19), no cutoff or tapering is applied in the calculation of $\chi^{(2)}_{ijk}(z, \omega_{IR})$, instead, the intramolecular part of $\chi^{(2)}_{ijk}(z, \omega_{IR})$ is smoothed with a Hanning-window of length $\omega^{intra}_{Hann} = 1.6$ THz and the intermolecular one with a broader window length of $\omega^{inter}_{Hann} = 14$ THz. Furthermore, $\chi^{(2)}_{ijk}(z, \omega_{IR})$ is smoothed in space with a Gaussian window function with a standard deviation of σ = 0.038 nm. To generate the necessary amount of data, we generate 90 starting configurations from a 18 ns long simulation with the SPC/E (20) force field using GROMACS (21) with a time step of 2 fs in the NVT ensemble, implemented by the CSVR-thermostat (22) with a relaxation time of 1 ps. For each of these 90 initial configurations we generate on average 0.24 ns long trajectories with the more expensive MB-Pol potential, using LAMMPS (14). In our analysis we discard the first 20 ps of each trajectory to give the system time to equilibrate to the new force field. Here we also use the CSVR-thermostat, but with a larger relaxation time of 5 ps and a smaller timestep of 0.2 fs. We simulate 352 water molecules in a box with the dimensions $L_x = L_y = 2$ nm in the plane of the interface and $L_z = 6$ nm orthogonal to it. The slab thickness determined by the distances between the two Gibbs-dividing surfaces is 2.64 nm. Electrostatic interactions are computed using periodic boundary conditions with the particle mesh Ewald method, the electric field along the z-axis is corrected for periodicity effects (8, 23). For the calculation of the effective polarizabilities a self-written Ewald-summation code is used.

## Theory Behind Depth-Resolved SFG/DFG Spectroscopy

An in-depth description of the theory underlying the depth-resolved SFG technique can be found in references (24, 25), only considering electric dipolar contributions. Here, the fundamental concepts of how it can be used to gain depth information are summarised and the theory is expanded to include electric quadrupolar sources in the following sections.

The measured SFG and DFG responses are governed by the effective second-order susceptibility, $\chi^{(2)}_{eff}$, which represents the spatially integrated response over the entire depth of the sample. For a signal sourcing from increasing depths, however, the input and output beams have longer path lengths and thus a propagation phase shift is introduced, $\phi_P$. This phase is given by the product of depth and the z-component of the wavevector mismatch, $\Delta k_z$, as in Eq. 10.

$$\phi_P = \Delta k_z z \tag{10}$$

Clearly, the phase of the output signal for chromophores at increasing depths continuously cycles around the full $2\pi$ range. When these signals are integrated, however, the coherence length, $1/\Delta k_z$, defines the amplitude scaling and the output response is phase-shifted by a 90° propagation phase, as in Eq. 11 (on the assumption that the generation of output signals tends to zero as the depth tends to infinity).

$$\int_0^\infty dz\, \chi^{(2)} e^{i\Delta k_z z} = \left[\frac{1}{i\Delta k_z} \chi^{(2)} e^{i\Delta k_z z}\right]_0^\infty$$

$$= \frac{i}{\Delta k_z} \chi^{(2)} \tag{11}$$

The above expression in Eq. 11, however, assumes that there is no depth dependency to $\chi^{(2)}$. This assumption is thus not generally valid for dipolar sources since only anisotropic regions yield any signals and most bulk media are isotropic (all bulk liquids). Therefore, $\chi^{(2)}$ is generally bound to the interfacial region and can be described by a depth-dependent decay function with decay length z'. With this depth-dependence included, the integration yields an effective susceptibility as in Eq. 12.

$$\chi^{(2)}_{eff} = \int_0^\infty dz\, \chi^{(2)} e^{-\frac{z}{z'}} e^{i\Delta k_z z}$$

$$= \frac{1}{\sqrt{\left(\frac{1}{z'}\right)^2 + (\Delta k_z)^2}} \chi^{(2)} e^{i\,\mathrm{atan}(\Delta k_z z')} \quad (12)$$

Clearly, the amplitude pre-factor scales as z' for short decay lengths, and tends to the coherence length as z' becomes large. Similarly, the phase-shift starts off approximately linear with z' but tends to ±90°, depending on the sign of $\Delta k_z$.

When probing resonances, the added propagation phase convolutes with the phase of the resonance, $\phi_R$. Therefore, as the resonant phase is typically unknown, the propagation phase is generally inseparable when measuring the phase-resolved spectra from a single second-order response, as indicated in Eq. 13.

$$\phi_{eff} = \phi_R + \phi_P \quad (13)$$

Similarly, as the amplitudes of the resonances (magnitude of $\chi^{(2)}$) are also not generally known, accessing any depth information from the amplitude ratio is also not possible.

By measuring both SFG and DFG phase-resolved responses, however, this convolution of resonant and depth information becomes separable. This is clearly the case as SFG and DFG involve different mixings of the input frequencies and thus must have different magnitude wavevector mismatches. For the input beams in this work, these values are ~0.02 nm$^{-1}$ for SFG and ~0.014 nm$^{-1}$ for DFG. It is important, however, to realise that the phase-resolved responses contain two halves of a complex conjugate pairing with negated frequency arguments, as described in Eq. 14 for SFG and Eq. 15 for DFG.

$$\chi^{(2)}_{eff}(\omega_2 + \omega_1, \omega_2, \omega_1) \quad vs. \quad \chi^{(2)}_{eff}(-\omega_2 - \omega_1, -\omega_2, -\omega_1) \quad (14)$$

$$\chi^{(2)}_{eff}(-\omega_2 + \omega_1, -\omega_2, \omega_1) \quad vs. \quad \chi^{(2)}_{eff}(\omega_2 - \omega_1, \omega_2, -\omega_1) \quad (15)$$

The sign of the resonant phase is dictated by the sign of the resonant frequency argument, taken to be $\omega_1$ with no loss in generality. Therefore, by considering half of each phase-resolved response with the same sign of $\omega_1$ (e.g., positive) the resonant information in each response is precisely equal. On the other hand, the sign of the wavevector mismatch is predominantly dictated by the signs of the two higher frequency arguments, which can clearly be seen to differ for the same sign of $\omega_1$. This makes the effective phases of each response as in Eqs. 16 and 17.

$$\phi^{SFG}_{eff} = \phi_R + |\phi^{SFG}_P| \quad (16)$$

$$\phi^{DFG}_{eff} = \phi_R - |\phi^{DFG}_P| \quad (17)$$

Thus, by taking the phase difference of the two responses, the depth information becomes accessible, as in Eq. 18.

$$\Delta\phi = \phi_{eff}^{SFG} - \phi_{eff}^{DFG} = \text{atan}(|\Delta k_z^{SFG}|z') + \text{atan}(|\Delta k_z^{DFG}|z') \qquad (18)$$

From this, it is clear that the sensitivity to depth predominantly comes from the difference in sign of the two wavevector mismatches, not their difference in amplitude. This is clear as the phase difference yields an approximately linear increase over the first 10 nm range of decay lengths of ~1.9°nm$^{-1}$, only ~0.3°nm$^{-1}$ of which comes from the magnitude mismatch. It is also worth noting the insensitivity of the amplitude ratio to decay length. As mentioned above, for relatively short decays, the amplitude pre-factor effectively scales as z' and is thus largely independent on $\Delta k_z$. Clearly, therefore, the amplitude ratio between DFG and SFG is effectively 1 for relatively short decay lengths. In fact, with the values for the wavevector mismatches here, it only yields ~1.01 for a 10 nm decay length, showing that it is effectively constant for the nanoscale decay lengths that could be expected for the structural anisotropy at an uncharged interface.

### *Inclusion of an Isotropic Contribution*

In the above treatment, it was assumed that only electric dipolar sources contribute, thus limiting the integration to the structurally anisotropic region in proximity to the interface. As shown in the later sections, however, quadrupolar sources are not restricted to anisotropic environments and thus can source throughout the entire bulk. For this contribution, the same integration behaviour as in Eq. 11 can be applied since it sources from the isotropic environments and thus effectivity has infinite depth and a depth-independent susceptibility. This isotropic contribution, however, is inherently phase shifted by 90° and multiplied by the wavevector of the upconversion beam (assuming SSP polarisation) due to its quadrupolar origin (see later). The overall response for SFG/DFG can thus be written as the sum of a decaying anisotropic contribution and a constant isotropic term, as in Eq. 19.

$$\chi_{eff}^{(2)SFG,DFG} = \int_0^\infty dz\, \chi_{aniso}^{(2)} e^{-\frac{z}{z'}} e^{i\Delta k_z^{SFG,DFG} z} + \int_0^\infty dz\, ik_2^{SFG,DFG} \chi_{iso}^{(2)} e^{i\Delta k_z^{SFG,DFG} z}$$

$$= \frac{1}{\sqrt{\left(\frac{1}{z'}\right)^2 + \left(\Delta k_z^{SFG,DFG}\right)^2}} \chi_{aniso}^{(2)} e^{i\,\text{atan}\left(\Delta k_z^{SFG,DFG} z'\right)} - \frac{k_2^{SFG,DFG}}{\Delta k_z^{SFG,DFG}} \chi_{iso}^{(2)} \qquad (19)$$

From this, it is clear that the anisotropic contribution yields a phase difference between SFG and DFG but essentially no amplitude difference (assuming a decay length <10 nm). In contrast, the isotropic contribution yields no phase difference but clearly different amplitudes. This is evident since, whilst SFG and DFG use the same upconversion field, the responses with equal resonant phases (i.e., positive $\omega_1$, see Eqs. 14 and 15) use different halves of its complex conjugate pairing i.e., have different signs for $k_2$. This sign difference is equally reflected in their wavevector mismatches and thus the only difference between the isotropic contribution for the two responses is from their respective coherence lengths (i.e., from the magnitude difference of their wavevector mismatches). Therefore, by combining the two contributions, namely anisotropic and isotropic, the amplitude ratio of the effective responses will solely speak to the relative proportion of the isotropic component, and the phase difference will be approximately linear with decay length (over a 10 nm range), but with a gradient that also depends on the isotropic proportion. This dependency is shown in Figure 1(d) in the main text. Through measuring both responses, each of these parameters can thus be determined.

Even for systems with longer anisotropic decay lengths, beyond which the assumption of equal amplitudes becomes poor, the measurement of both SFG and DFG still allows for a full determination of the relative isotropic proportion and decay length of the anisotropy. This is clear as the phase difference and amplitude ratio for the anisotropic contribution are inherently linked and thus should both reflect the exact decay length. A modelled plot of both parameters is shown in Figure S1 for anisotropic decay lengths up to 100 nm and relative isotropic contributions (ratios to the anisotropic component) varying from +1.0 to -1.0. From this, it is clear that an experimentally determined pair of values for the two parameters can be used to directly access both quantities, with an example being highlighted for an observed phase difference of 60° and an amplitude ratio of 1.1.

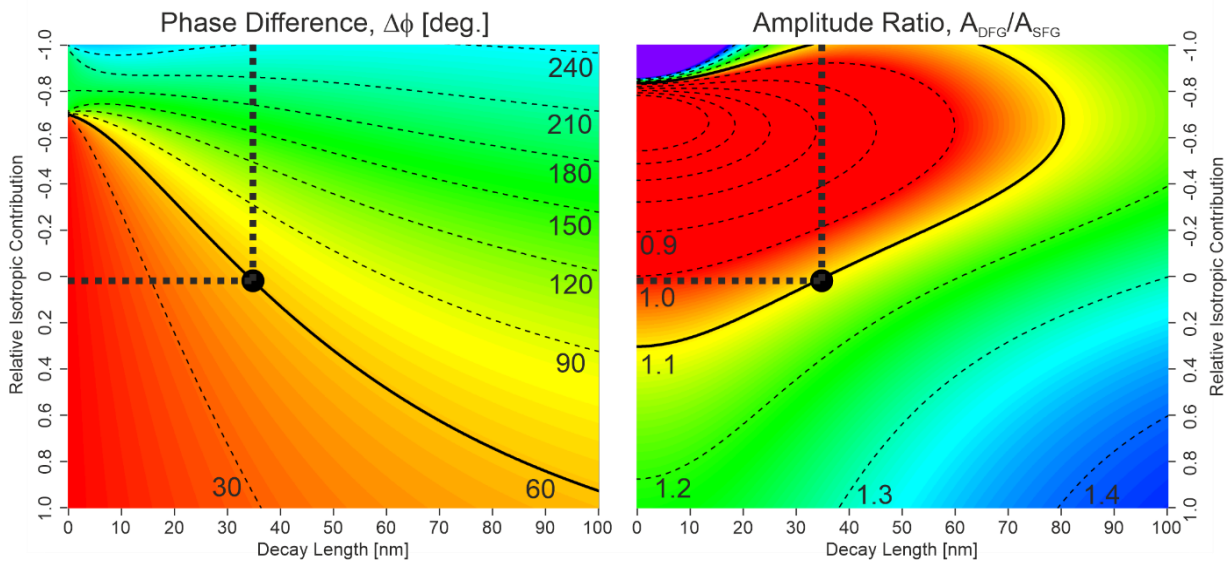

*Figure S1: Modelled plots of the SFG-DFG phase difference and amplitude ratio as a function of both the anisotropic decay length and the relative isotropic contribution. Contour lines are also included for better clarity and an example extraction of the desired quantities is shown for a phase difference of 60° and amplitude ratio of 1.1.*

## Absolute zero positions

In the above theoretical description of the depth-resolved SFG-DFG spectroscopy it is clear that the propagation phase can be extracted, but it is not immediately obvious what the z=0 position is which yields no additional phase. However, given that the signals generated from a sample surface are heterodyned with SFG and DFG local oscillators, these reference signals provide the phase reference for each pulse. As such, since they are linearly reflected from the same surface, the reference depth must be the plane of linear reflection (PLR). This effective depth is hence governed by the gradient in the dielectric function, and thus exists very close to the top-most surface region.

As noted above, however, the absolute phases are initially determined by referencing to a measurement from z-cut quartz, assuming it to be an entirely bulk dipolar signal, thus with a phase of ±90°. However, since quartz is known to deviate slightly from this phase owing to its structural changes at the surface(4), the absolute phases are then corrected by a SFG-DFG measurement from a self-assembled monolayer of OTS on fused silica. Specifically, the methyl resonances from the OTS film are phase-rotated such that they are exactly equal in SFG and

DFG. This hence sets the zero-depth position to be the position of the methyl groups, not the PLR i.e., z=0 is effectively set to the on-set of density change from air into the sample. Therefore, given that the change in both density and dielectric function is relatively similar across the air-water and air-SAM-FS interfaces, this depth correction factor is expected to also essentially move the z=0 position from the PLR to the top-most water layer which corresponds to the on-set of the density change.

For the simulation results, the boundary plane is instead defined as the Gibbs diving surface (GDS) which represents the midpoint of the density change between the two phases. Therefore, the simulated depths will be offset to the experimental depth by approximately half the width of the density change profile. As the density changes over ~3-5Å, the simulated depths are thus expected to be offset by about -2Å. By comparing the data in Figure 4 in the main text, it is clear that the simulations do indeed predict the on-set of the SFG signals to occur at ~-2Å.

## Quadrupolar Contributions to the Water Response

From the spectra shown in Figure 2(a) in the main text, there is a clear offset in amplitude between SFG and DFG in the real part of the measured response, but no distinct difference in imaginary part. Within the electric dipole approximation, the intrinsic SFG and DFG responses are equal in the absence of dispersion (as shown by the permutation symmetry relations discussed below). Since dispersion is shown to be insignificant in Figure 3(b), any discrepancy between SFG and DFG can only arise from the integration of the (potentially depth-dependent) intrinsic responses. Whilst a depth in the anisotropy does lead to a larger DFG response cf. SFG, it also introduces a phase difference between SFG and DFG for dipolar sources.(24, 25) However, the spectra in Figure 1(a) clearly show that there is negligable phase difference between SFG and DFG, indicating that the source of this discrepancy cannot arise from depth based on the electric dipole mechanism. Instead, however, one must look beyond the electric dipole contributions to explain this observation.

In the following discussion, the second-order responses are expanded beyond the electric dipole contribution to include quadrupolar mechanisms, broadly following the theoretical treatment and notation given by Morita(5). Initially, the terms arising from these mechanisms are derived and their contributions to SFG and DFG pathways compared. In this context, there are two important scenarios to consider. The first is the comparison between SFG and DFG responses produced simultaneously from the same experiment and thus using the same input beams (but therefore having different output frequencies). In this scenario, there is no inherent permutation symmetry in the intrinsic dipolar susceptibilities for both outputs. The second scenario compares the SFG and DFG responses involving the same three beams, where the DFG is produced by using an upconversion beam at the SFG frequency (labelled DFG' as in the main text). Whilst permutation symmetry for the dipolar response hold in this comparison, the two responses must necessarily arise from separate measurements.(24)

Once the expressions behind these comparisons are derived in each scenario, we then move to analyse the measured non-resonant response from $H_2O$. From the amplitude ratios between the SFG and both DFG responses, it is determined that the bulk quadrupolar contribution appears to be the source of the observed discrepancy. Thereafter, the implications of these results are discussed.

*Expansion to the Quadrupolar Terms*

By analogy to the linear polarisability, the second-order susceptibility can be expanded in a Taylor series based on its spatial coordinates, as in Eq. 20.

$$\chi^{(2)} = \chi_0^{(2)} + \left(\frac{\partial \chi^{(2)}}{\partial r}\right)_{r_0} (r - r_0) + \cdots \quad (20)$$

This expansion can then be input into the expression for the second-order polarisation, as in Eq. 21.

$$P^{(2)} = \varepsilon_0 \chi^{(2)} : E_1 E_2$$
$$= \varepsilon_0 \chi_0^{(2)} : E_1 E_2 + \varepsilon_0 \left(\frac{\partial \chi^{(2)}}{\partial r}\right)_{r_0} (r - r_0) : E_1 E_2 + \cdots \quad (21)$$

Therefore, by means of the product rule, the second term can be rewritten as in Eq. 22.

$$\left(\frac{\partial \chi^{(2)}}{\partial r}\right)_{r_0} (r - r_0) : E_1 E_2 = \frac{\partial}{\partial r}\left(\chi^{(2)}(r - r_0) : E_1 E_2\right) - \chi^{(2)} : E_1 E_2$$
$$- \chi^{(2)}(r - r_0) : \frac{\partial E_1}{\partial r} E_2 - \chi^{(2)}(r - r_0) : E_1 \frac{\partial E_2}{\partial r} \quad (22)$$

The four terms on the right side of Eq. 22 each describe different mechanistic pathways towards generation of a second-order polarisation. Specifically, the first term, involving the derivative of the whole product, describes the dipole produced from the gradient of an induced quadrupole. By contrast, the second term simply describes the dipole produced by the electric dipole mechanism. Finally, the remaining two terms describe dipoles produced using both an electric field and the gradient of another electric field, and thus must be quadrupolar in origin (neglecting magnetic dipole mechanisms). Clearly, the three terms based on quadrupolar processes are described by a rank four tensors, and are thus non-cancelling in isotropic environments. Based on this description of the different processes, the four terms can be written with unique susceptibilities labelled as Q, D0, D1, and D2, respectively, indicating the mechanism of generation. The second-order polarisation can thus be written as in Eq. 23, assuming no inherent polarisation.

$$P^{(2)} = \varepsilon_0 \chi^{(2),D0} : E_1 E_2 + \varepsilon_0 \chi^{(2),D1} : \frac{\partial E_1}{\partial r} E_2 + \varepsilon_0 \chi^{(2),D2} : E_1 \frac{\partial E_2}{\partial r} - \varepsilon_0 \frac{\partial}{\partial r}\left(\chi^{(2),Q} : E_1 E_2\right) \quad (23)$$

In order to describe the full measured response, one must then integrate these contributions over infinite depth through the sample. The effective second-order susceptibility is then given by Eq. 24, where the generation processes are driven using local fields, $\tilde{E}_i^n$, which account for both Fresnel factors, $L_{ii}^n$, and depth-dependent local field factors, $f_i^n$, as in Eq. 25.

$$\chi_{ijk}^{(2),eff} E_j^2 E_k^1 = \sum_{jks}^{x,y,z} \int_{-\infty}^{\infty} dz\, f_i^3 L_{ii}^3 \Big[ \chi_{ijk}^{(2),D0}(\omega_3,\omega_2,\omega_1,z) f_k^1 L_{kk}^1 E_k^1 f_j^2 L_{jj}^2 E_j^2$$

$$+ \chi_{ijks}^{(2),D1}(\omega_3,\omega_2,\omega_1,z) \frac{\partial f_k^1 L_{kk}^1 E_k^1}{\partial s} f_j^2 L_{jj}^2 E_j^2$$

$$+ \chi_{ijks}^{(2),D2}(\omega_3,\omega_2,\omega_1,z) f_k^1 L_{kk}^1 E_k^1 \frac{\partial f_j^2 L_{jj}^2 E_j^2}{\partial s}$$

$$- \frac{\partial}{\partial s}\Big( \chi_{ijks}^{(2),Q}(\omega_3,\omega_2,\omega_1,z) f_k^1 L_{kk}^1 E_k^1 f_j^2 L_{jj}^2 E_j^2 \Big) \Big] e^{-k_z^3 z} \qquad (24)$$

$$\tilde{E}_i^n(z) = f_i^n(z) L_{ii}^n E_i^n \qquad (25)$$

The integral in Eq. 24 may then be broken down based on knowledge about the spatial variation and symmetry properties of the different parameters. Firstly, the Fresnel factors are spatially invariant and purely related to the beam polarisation, experimental geometry, and dielectric constants (which give them frequency dependence). By contrast, both the local field factors and the input fields have spatial variations. The local field factors only vary with depth very close to the interface due to the continuous change in the dielectric function between the incident and bulk media. On the other hand, the incident fields only have spatial variations due to their oscillatory nature, thus depending on their wavevectors, and exist throughout the entire integral range. This allows the integral to be broken up into different contributions, namely those isolated to the interfacial range ($z = 0 \to z_0$) and those covering the entire sample ($z = 0 \to \infty$), as in Eq. 26.

$$\chi_{ijk}^{(2),eff} E_j^2 E_k^1 = \sum_{jks}^{x,y,z} L_{ii}^3 L_{jj}^2 L_{kk}^1 \Bigg[ \int_0^{z_0} dz\, \chi_{ijk}^{(2),D0}(\omega_3,\omega_2,\omega_1,z) f_k^1(z) f_j^2(z) f_i^3(z) E_k^1 E_j^2 e^{-ik_z^3 z} \quad (26a)$$

$$+ \int_0^{z_0} dz\, \chi_{ijkz}^{(2),D1}(\omega_3,\omega_2,\omega_1,z) \frac{\partial f_k^1(z)}{\partial z} f_j^2(z) f_i^3(z) E_k^1 E_j^2 e^{-ik_z^3 z} \qquad (26b)$$

$$+ \int_0^{\infty} dz\, \chi_{ijks}^{(2),D1}(\omega_3,\omega_2,\omega_1,z) f_k^1(z) f_j^2(z) f_i^3(z) \frac{\partial E_k^1}{\partial s} E_j^2 e^{-ik_z^3 z} \qquad (26c)$$

$$+ \int_0^{z_0} dz\, \chi_{ijkz}^{(2),D2}(\omega_3,\omega_2,\omega_1,z) f_k^1(z) \frac{\partial f_j^2(z)}{\partial z} f_i^3(z) E_k^1 E_j^2 e^{-ik_z^3 z} \qquad (26d)$$

$$+ \int_0^{\infty} dz\, \chi_{ijks}^{(2),D2}(\omega_3,\omega_2,\omega_1,z) f_k^1(z) f_j^2(z) f_i^3(z) E_k^1 \frac{\partial E_j^2}{\partial s} e^{-ik_z^3 z} \qquad (26e)$$

$$- \int_0^{z_0} dz\, f_i^3(z) \frac{\partial}{\partial z}\Big( \chi_{ijkz}^{(2),Q}(\omega_3,\omega_2,\omega_1,z) f_k^1(z) f_j^2(z) \Big) E_k^1 E_j^2 e^{-ik_z^3 z} \qquad (26f)$$

$$- \int_0^{\infty} dz\, \chi_{ijks}^{(2),Q}(\omega_3,\omega_2,\omega_1,z) f_k^1(z) f_j^2(z) f_i^3(z) \frac{\partial}{\partial s}\Big( E_k^1 E_j^2 \Big) e^{-ik_z^3 z} \Bigg] \qquad (26g)$$

For the terms involving integration over the full depth (26c, e, and g), any variation in both the local field factors and susceptibilities can be neglected as the vast majority of the signal arises from the isotropic bulk. Since the $z_0$ depth is defined to be the point at which any anisotropy vanishes, the bulk quantities can be taken as those evaluated at this depth. On the other hand, for the interfacial terms, any variation in the oscillatory field contributions can also be neglected on the assumption that $z_0$ is much smaller than the wavelength. This allows Eq. 26 to be simplified as in Eq. 27, having explicitly included the derivatives of the oscillatory fields in the bulk terms.

$$\chi_{ijk}^{(2),eff} E_j^2 E_k^1 = \sum_{jks}^{x,y,z} L_{ii}^3 L_{jj}^2 L_{kk}^1 \left[ \int_0^{z_0} dz \left( \chi_{ijk}^{(2),D0}(\omega_3,\omega_2,\omega_1,z) f_k^1(z) f_j^2(z) f_i^3(z) \right) \right. \quad (27a)$$

$$+ \int_0^{z_0} dz \, \chi_{ijkz}^{(2),D1}(\omega_3,\omega_2,\omega_1,z) \frac{\partial f_k^1(z)}{\partial z} f_j^2(z) f_i^3(z) \quad (27b)$$

$$+ \chi_{ijks}^{(2),D1}(\omega_3,\omega_2,\omega_1,z_0) f_k^1(z_0) f_j^2(z_0) f_i^3(z_0) \int_0^{\infty} dz \, i k_s^1 e^{i(k_z^1+k_z^2-k_z^3)z} \quad (27c)$$

$$+ \int_0^{z_0} dz \, \chi_{ijkz}^{(2),D2}(\omega_3,\omega_2,\omega_1,z) f_k^1(z) \frac{\partial f_j^2(z)}{\partial z} f_i^3(z) \quad (27d)$$

$$+ \chi_{ijks}^{(2),D2}(\omega_3,\omega_2,\omega_1,z_0) f_k^1(z_0) f_j^2(z_0) f_i^3(z_0) \int_0^{\infty} dz \, i k_s^2 e^{i(k_z^1+k_z^2-k_z^3)z} \quad (27e)$$

$$- \int_0^{z_0} dz \, f_i^3(z) \frac{\partial}{\partial z} \left( \chi_{ijkz}^{(2),Q}(\omega_3,\omega_2,\omega_1,z) f_k^1(z) f_j^2(z) \right) \quad (27f)$$

$$\left. - \chi_{ijks}^{(2),Q}(\omega_3,\omega_2,\omega_1,z_0) f_k^1(z_0) f_j^2(z_0) f_i^3(z_0) \int_0^{\infty} dz \, i(k_s^1+k_s^2) e^{i(k_z^1+k_z^2-k_z^3)z} \right] |E_k^1||E_j^2| \quad (27g)$$

The three bulk terms can then be directly evaluated, yielding the terms 28c, e, and g. Similarly, by integration by parts, term 27f can be expanded into two contributions, as terms 28f.1 and 28f.2 in Eq. 28.

$$\chi_{ijk}^{(2),eff} E_j^2 E_k^1 = \sum_{jks}^{x,y,z} L_{ii}^3 L_{jj}^2 L_{kk}^1 \left[ \int_0^{z_0} dz \left( \chi_{ijk}^{(2),D0}(\omega_3,\omega_2,\omega_1,z) f_k^1(z) f_j^2(z) f_i^3(z) \right) \right. \quad (28a)$$

$$+ \int_0^{z_0} dz \, \chi_{ijkz}^{(2),D1}(\omega_3,\omega_2,\omega_1,z) \frac{\partial f_k^1(z)}{\partial z} f_j^2(z) f_i^3(z) \quad (28b)$$

$$- \chi_{ijks}^{(2),D1}(\omega_3,\omega_2,\omega_1,z_0) f_k^1(z_0) f_j^2(z_0) f_i^3(z_0) \frac{k_s^1}{k_z^1+k_z^2-k_z^3} \quad (28c)$$

$$+ \int_0^{z_0} dz\, \chi^{(2),D2}_{ijkz}(\omega_3, \omega_2, \omega_1, z) f_k^1(z) \frac{\partial f_j^2(z)}{\partial z} f_i^3(z) \tag{28d}$$

$$-\chi^{(2),D2}_{ijks}(\omega_3, \omega_2, \omega_1, z_0) f_k^1(z_0) f_j^2(z_0) f_i^3(z_0) \frac{k_s^2}{k_z^1 + k_z^2 - k_z^3} \tag{28e}$$

$$+ \int_0^{z_0} dz\, \frac{\partial f_i^3(z)}{\partial z} \chi^{(2),Q}_{ijkz}(\omega_3, \omega_2, \omega_1, z) f_k^1(z) f_j^2(z) \tag{28f.1}$$

$$-\chi^{(2),Q}_{ijkz}(\omega_3, \omega_2, \omega_1, z_0) f_k^1(z_0) f_j^2(z_0) f_i^3(z_0) \tag{28f.2}$$

$$+\chi^{(2),Q}_{ijks}(\omega_3, \omega_2, \omega_1, z_0) f_k^1(z_0) f_j^2(z_0) f_i^3(z_0) \frac{k_s^1 + k_s^2}{k_z^1 + k_z^2 - k_z^3} \Bigg] |E_k^1||E_j^2| \tag{28g}$$

These terms can be grouped together based on the mechanism of their dipole induction and source location, with 28a being the interfacial dipole term (ID), 28b, 28d, and 28f.1 being interfacial quadrupole terms (IQ), 28c, 28e, and 28g being bulk quadrupole terms (QB), and 28f.2 being an interfacial quadrupolar contribution but with bulk-like properties (IQB). The effective susceptibility can thus be written as a sum of these contributions, as in Eq. 29.

$$\chi^{(2)}_{eff} = \chi^{(2)ID} + \chi^{(2)IQ} + \chi^{(2)IQB} + \chi^{(2)QB} \tag{29}$$

*Comparing SFG and DFG Responses*
If we start by comparing the SFG and DFG responses that involve the same three frequencies (SFG and DFG'), and specifically only talking one half of the phase-resolved response involving a positive $\omega_1$ argument, one must consider the permutation symmetry between the intrinsic susceptibilities. By limiting our considerations to processes that start in the ground vibrational state, then the SFG and DFG' processes are contributed to by two and three pairs of correlation functions (pathways), with each pair being complex conjugates that just represent a frequency negation.(24) Specifically, by taking only positive $\omega_1$ as above, the SFG and DFG' susceptibilities can be written as in Eqs. 30 and 31, where $a$, $b$, and $c$ are used in the numerators to represent either $\mu$ or $q$ (where the additional tensor dimension, s, is associated only with $q$). This allows us to consider either dipolar or quadrupolar transitions under any of the aforementioned mechanisms. (Note that the DFG' response deliberately has the '$i$' and '$j$' indices reversed.)

$$\tilde{\chi}^{(2)}_{ijk(s)}(\omega_2 + \omega_1, \omega_2, \omega_1) = \frac{N}{2\varepsilon_0 \hbar^2} \sum_{qr} \frac{a^{(s)i}_{gr} b^{(s)j}_{rq} c^{(s)k}_{qg}}{(\omega_1 - \omega_{qg} + i\Gamma_{qg})(\omega_2 + \omega_1 - \omega_{rg} + i\Gamma_{rg})} \tag{30.1}$$

$$+ \frac{a^{(s)i}_{gr} b^{(s)j}_{qg} c^{(s)k}_{rq}}{(\omega_2 - \omega_{qg} + i\Gamma_{qg})(\omega_2 + \omega_1 - \omega_{rg} + i\Gamma_{rg})} \tag{30.2}$$

$$\tilde{\chi}^{(2)}_{jik(s)}(-\omega_2, -\omega_2 - \omega_1, \omega_1) = \frac{N}{2\varepsilon_0 \hbar^2} \sum_{qr} \frac{a^{(s)i}_{gr} b^{(s)j}_{rq} c^{(s)k}_{qg}}{(\omega_1 - \omega_{qg} + i\Gamma_{qg})(\omega_2 - \omega_{rq} - i\Gamma_{rq})} \tag{31.1}$$

$$-\frac{a_{gr}^{(s)i} b_{rq}^{(s)j} c_{qg}^{(s)k}}{(\omega_2 + \omega_1 - \omega_{rg} - i\Gamma_{rg})(\omega_2 - \omega_{rq} - i\Gamma_{rq})} \tag{31.2}$$

$$+\frac{a_{gr}^{(s)i} b_{qg}^{(s)j} c_{rq}^{(s)k}}{(\omega_2 + \omega_1 - \omega_{rg} - i\Gamma_{rg})(\omega_2 - \omega_{qg} - i\Gamma_{qg})} \tag{31.3}$$

The DFG' terms 31.1 and 31.2 can be trivially factorised and combined to yield Eq. 32.

$$\tilde{\chi}_{jik(s)}^{(2)}(-\omega_2, -\omega_2 - \omega_1, \omega_1) =$$

$$\frac{N}{2\varepsilon_0 \hbar^2} \sum_{qr} \frac{a_{gr}^{(s)i} b_{rq}^{(s)j} c_{qg}^{(s)k}}{(\omega_2 - \omega_{rq} - i\Gamma_{rq})} \left\{ \frac{\omega_2 - \omega_{rq} - i(\Gamma_{rg} + \Gamma_{qg})}{(\omega_1 - \omega_{qg} + i\Gamma_{qg})(\omega_2 + \omega_1 - \omega_{rg} - i\Gamma_{rg})} \right\} \tag{32.1}$$

$$+\frac{a_{gr}^{(s)i} b_{qg}^{(s)j} c_{rq}^{(s)k}}{(\omega_2 + \omega_1 - \omega_{rg} - i\Gamma_{rg})(\omega_2 - \omega_{qg} - i\Gamma_{qg})} \tag{32.2}$$

Whilst this is not the same as the analogous expression for SFG in Eq. 30, when the two higher frequency arguments are off-resonant, the two expressions clearly converge, independent on the transition mechanisms. This shows that the permutation symmetry holds for all of the aforementioned intrinsic susceptibilities (D0, D1, D2, and Q) as long as $i = j$, which is necessarily the case when using the SSP polarisation combination, as used here.

*Interfacial Dipole Contribution (ID)*

Firstly, the dipole responses from SFG and DFG' are considered, being given in Eqs. 33 and 34, respectively, exploiting the intrinsic permutation symmetry relation between the two susceptibilities along with the symmetry of the local field factors on negation of the input frequencies. Clearly, the overall SFG and DFG' responses from this mechanism are equal.

$$\chi_{SSP}^{(2)ID}(\omega_2 + \omega_1, \omega_2, \omega_1) =$$

$$\sum_{k}^{x,z} \int_0^{z_0} dz\, \chi_{yyk}^{(2),D0}(\omega_2 + \omega_1, \omega_2, \omega_1, z) f_k(\omega_1, z) f_y(\omega_2, z) f_y(\omega_2 + \omega_1, z) \tag{33}$$

$$\chi_{SSP}^{(2)ID}(-\omega_2, -\omega_2 - \omega_1, \omega_1) =$$

$$\sum_{k}^{x,z} \int_0^{z_0} dz\, \chi_{yyk}^{(2),D0}(\omega_2 + \omega_1, \omega_2, \omega_1, z) f_k(\omega_1, z) f_y(\omega_2, z) f_y(\omega_2 + \omega_1, z) \tag{34}$$

*Interfacial Quadrupolar Contribution (IQ)*

Next, if we consider the interfacial quadrupolar contributions to the same two responses, they are given by Eqs. 35 and 36.

$$\chi_{SSP}^{(2)IQ}(\omega_2 + \omega_1, \omega_2, \omega_1)$$
$$= \sum_{k}^{x,z} \int_0^{z_0} dz \left[ \chi_{yykz}^{(2),D1}(\omega_2 + \omega_1, \omega_2, \omega_1, z) \frac{\partial f_k(\omega_1, z)}{\partial z} f_y(\omega_2, z) f_y(\omega_2 + \omega_1, z) \right.$$
$$\left. + \chi_{yykz}^{(2),D2}(\omega_2 + \omega_1, \omega_2, \omega_1, z) f_k(\omega_1, z) \frac{\partial f_y(\omega_2, z)}{\partial z} f_y(\omega_2 + \omega_1, z) \right.$$

$$+ \chi^{(2),Q}_{yykz}(\omega_2 + \omega_1, \omega_2, \omega_1, z) f_k(\omega_1, z) f_y(\omega_2, z) \frac{\partial f_y(\omega_2 + \omega_1, z)}{\partial z} \Bigg] \quad (35)$$

$$\chi^{(2)IQ}_{SSP}(-\omega_2, -\omega_2 - \omega_1, \omega_1)$$
$$= \sum_k^{x,z} \int_0^{z_0} dz \Bigg[ \chi^{(2),D1}_{yykz}(\omega_2 + \omega_1, \omega_2, \omega_1, z) \frac{\partial f_k(\omega_1, z)}{\partial z} f_y(\omega_2, z) f_y(\omega_2 + \omega_1, z)$$
$$+ \chi^{(2),D2}_{yykz}(\omega_2 + \omega_1, \omega_2, \omega_1, z) f_k(\omega_1, z) \frac{\partial f_y(\omega_2 + \omega_1, z)}{\partial z} f_y(\omega_2, z)$$
$$+ \chi^{(2),Q}_{yykz}(\omega_2 + \omega_1, \omega_2, \omega_1, z) f_k(\omega_1, z) f_y(\omega_2 + \omega_1, z) \frac{\partial f_y(\omega_2, z)}{\partial z} \Bigg] \quad (36)$$

At this point, however, it is clear that the interfacial dipole responses for SFG and DFG' are not intrinsically the same. Whilst the D1 contribution is equivalent to both, the D2 and Q terms have swapped their local field factor modulation functions, with these not generally being equal, as in Eq. 37.

$$\frac{\partial f_y(\omega_2 + \omega_1, z)}{\partial z} f_y(\omega_2, z) \neq f_y(\omega_2 + \omega_1, z) \frac{\partial f_y(\omega_2, z)}{\partial z} \quad (37)$$

Nevertheless, as these two frequencies are off-resonant and typically close in frequency (with $\omega_1$ being in the mid-IR), any dispersion in their dielectric functions is likely small and thus, to a good approximation, the inequality in Eq. 37 can be taken to be equal.

By considering the simultaneously generated DFG response (with the same upconversion but different output frequency as the SFG), similar expressions can be generated only now without permutation symmetry between the intrinsic susceptibilities, as shown in Eq. 38.

$$\chi^{(2)IQ}_{SSP}(-\omega_2 + \omega_1, -\omega_2, \omega_1)$$
$$= \sum_k^{x,z} \int_0^{z_0} dz \Bigg[ \chi^{(2),D1}_{yykz}(-\omega_2 + \omega_1, -\omega_2, \omega_1, z) \frac{\partial f_k(\omega_1, z)}{\partial z} f_y(\omega_2, z) f_y(\omega_2 - \omega_1, z)$$
$$+ \chi^{(2),D2}_{yykz}(-\omega_2 + \omega_1, -\omega_2, \omega_1, z) f_k(\omega_1, z) \frac{\partial f_y(\omega_2, z)}{\partial z} f_y(\omega_2 - \omega_1, z)$$
$$+ \chi^{(2),Q}_{yykz}(-\omega_2 + \omega_1, -\omega_2, \omega_1, z) f_k(\omega_1, z) f_y(\omega_2, z) \frac{\partial f_y(\omega_2 - \omega_1, z)}{\partial z} \Bigg] \quad (38)$$

Now, equality between the SFG and DFG only exists on the assumption of no dispersion in the high frequency range, both in the local field factors and the intrinsic susceptibilities. Both this assumption and the one above for the permutation-symmetric response are, however, somewhat experimentally validated in Figure 3(b) in the main text where the two different DFG responses (DFG and DFG') are compared, showing only minute differences.

*Interfacial Quadrupolar Contribution with Bulk-like Properties (IQB)*

The comparison between SFG and DFG' for the interfacial quadrupolar contribution with bulk-like properties is very straightforward, as it only constitutes a single term with no derivatives of local field factors. The two contributions can be written as in Eqs. 39 and 40, again including the permutation symmetry for DFG'.

$$\chi^{(2)IQB}_{SSP}(\omega_2+\omega_1,\omega_2,\omega_1)=$$

$$-\sum_k^{x,z}\chi^{(2),Q}_{yykz}(\omega_2+\omega_1,\omega_2,\omega_1,z_0)f_k(\omega_1,z_0)f_y(\omega_2,z_0)f_y(\omega_2+\omega_1,z_0) \quad (39)$$

$$\chi^{(2)IQB}_{SSP}(-\omega_2,-\omega_2-\omega_1,\omega_1)=$$

$$-\sum_k^{x,z}\chi^{(2),Q}_{yykz}(\omega_2+\omega_1,\omega_2,\omega_1,z_0)f_k(\omega_1,z_0)f_y(\omega_2,z_0)f_y(\omega_2+\omega_1,z_0) \quad (40)$$

This shows that the two responses are identical. If we then consider the simultaneous DFG response, a similar expression to Eq. 40 can be generated, only with a frequency shift where equality is only reached with no dispersion.

*Bulk Quadrupolar Contribution (QB)*

For the bulk quadrupolar contributions, the SFG and permutation-symmetric DFG' responses are given by Eqs. 41 and 42.

$$\chi^{(2)QB}_{SSP}(\omega_2+\omega_1,\omega_2,\omega_1)=\sum_s^{x,y,z}\sum_k^{x,z}\frac{f_k(\omega_1,z_0)f_y(\omega_2,z_0)f_y(\omega_2+\omega_1,z_0)}{|k_z(\omega_1)|+|k_z(\omega_2)|+|k_z(\omega_1+\omega_2)|}$$

$$\left\{-|k_s(\omega_1)|\chi^{(2),D1}_{yyks}(\omega_2+\omega_1,\omega_2,\omega_1,z_0)-|k_s(\omega_2)|\chi^{(2),D2}_{yyks}(\omega_2+\omega_1,\omega_2,\omega_1,z_0)\right.$$

$$\left.+(|k_s(\omega_1)|+|k_s(\omega_2)|)\chi^{(2),Q}_{yyks}(\omega_2+\omega_1,\omega_2,\omega_1,z_0)\right\} \quad (41)$$

$$\chi^{(2)QB}_{SSP}(-\omega_2,-\omega_2-\omega_1,\omega_1)=\sum_s^{x,y,z}\sum_k^{x,z}\frac{f_k(\omega_1,z_0)f_y(\omega_2,z_0)f_y(\omega_2+\omega_1,z_0)}{|k_z(\omega_1)|-|k_z(\omega_2+\omega_1)|-|k_z(\omega_2)|}$$

$$\left\{-|k_s(\omega_1)|\chi^{(2),D1}_{yyks}(\omega_2+\omega_1,\omega_2,\omega_1,z_0)+|k_s(\omega_2+\omega_1)|\chi^{(2),D2}_{yyks}(\omega_2+\omega_1,\omega_2,\omega_1,z_0)\right.$$

$$\left.+(|k_s(\omega_1)|-|k_s(\omega_2+\omega_1)|)\chi^{(2),Q}_{yyks}(\omega_2+\omega_1,\omega_2,\omega_1,z_0)\right\} \quad (42)$$

These two expressions can be simplified by noting the transversality of fields, which means the wavevector and field must locally be orthogonal, as described by Eq. 43.

$$\sum_i^{x,y,z}k_i(\omega)f_i(\omega)L_{ii}(\omega)E_i(\omega)=\boldsymbol{k}(\omega)\cdot\boldsymbol{E}_{loc}(\omega)=0 \quad (43)$$

Based on the defined laboratory coordinates, the k-vectors are fixed within the xz-plane. Therefore, the summation over 's' is restricted to x and z, with the y-contributions vanishing. Furthermore, based on an isotropic symmetry argument, the only non-vanishing susceptibility terms are the YYXX and YYZZ, meaning $s=k$. Based on this, the D1 quadrupolar contribution must vanish along with the $k_s(\omega_1)$ contribution to the Q term. This means the SFG and DFG' expressions can be simplified to Eqs. 44 and 45.

$$\chi^{(2)QB}_{SSP}(\omega_2+\omega_1,\omega_2,\omega_1)=\sum_k^{x,z}\frac{f_k(\omega_1,z_0)f_y(\omega_2,z_0)f_y(\omega_2+\omega_1,z_0)}{|k_z(\omega_1)|+|k_z(\omega_2)|+|k_z(\omega_1+\omega_2)|}$$

$$|k_k(\omega_2)|\left\{-\chi^{(2),D2}_{yykk}(\omega_2+\omega_1,\omega_2,\omega_1,z_0)+\chi^{(2),Q}_{yykk}(\omega_2+\omega_1,\omega_2,\omega_1,z_0)\right\} \quad (44)$$

$$\chi_{SSP}^{(2)QB}(-\omega_2, -\omega_2 - \omega_1, \omega_1) = \sum_k^{x,z} \frac{f_k(\omega_1, z_0) f_y(\omega_2, z_0) f_y(\omega_2 + \omega_1, z_0)}{|k_z(\omega_1)| - |k_z(\omega_2 + \omega_1)| - |k_z(\omega_2)|}$$
$$|k_k(\omega_2 + \omega_1)| \left\{ \chi_{yykk}^{(2),D2}(\omega_2 + \omega_1, \omega_2, \omega_1, z_0) - \chi_{yykk}^{(2),Q}(\omega_2 + \omega_1, \omega_2, \omega_1, z_0) \right\} \quad (45)$$

Clearly, there is significant similarity between the two expressions, but they are modulated by different combinations of wavevectors. Specifically, the ratio of the DFG' and SFG responses is given in Eq. 46. It is worth noting that the k-subscripts of the k-vectors on the right side of Eq. 46 have been neglected since both contributions arise from the upconversion beam in each experiment and are thus assumed to have the same incidence angle and minimal dispersion.

$$\frac{\chi_{SSP}^{(2)QB}(-\omega_2, -\omega_2 - \omega_1, \omega_1)}{\chi_{SSP}^{(2)QB}(\omega_2 + \omega_1, \omega_2, \omega_1)} = \frac{|k_z(\omega_1 + \omega_2)| + |k_z(\omega_2)| + |k_z(\omega_1)|}{|k_z(\omega_2 + \omega_1)| + |k_z(\omega_2)| - |k_z(\omega_1)|} \frac{|k(\omega_2 + \omega_1)|}{|k(\omega_2)|} \quad (46)$$

If one ignores dispersion completely, also taking a collinear beam geometry as in used in the work presented here, then Eq. 46 becomes Eq. 47.

$$\frac{\chi_{SSP}^{(2)QB}(-\omega_2, -\omega_2 - \omega_1, \omega_1)}{\chi_{SSP}^{(2)QB}(\omega_2 + \omega_1, \omega_2, \omega_1)} \approx \left| \frac{k(\omega_2 + \omega_1)}{k(\omega_2)} \right|^2 \quad (47)$$

By considering the DFG response simultaneously generated with the SFG, a similar process yields Eq. 48.

$$\chi_{SSP}^{(2)QB}(-\omega_2 + \omega_1, -\omega_2, \omega_1) = \sum_k^{x,z} \frac{f_k(\omega_1, z_0) f_y(\omega_2, z_0) f_y(\omega_2 - \omega_1, z_0)}{|k_z(\omega_2 - \omega_1)| + |k_z(\omega_2)| - |k_z(\omega_1)|}$$
$$|k_k(\omega_2)| \left\{ -\chi_{yykk}^{(2),D2}(\omega_2, \omega_2 - \omega_1, \omega_1, z_0) + \chi_{yykk}^{(2),Q}(\omega_2, \omega_2 - \omega_1, \omega_1, z_0) \right\} \quad (48)$$

Therefore, by comparison to Eq. 44, the ratio between DFG and SFG is given by Eq. 49, assuming no dispersion. This is purely the ratio of their coherence lengths.

$$\frac{\chi_{SSP}^{(2)QB}(-\omega_2 + \omega_1, -\omega_2, \omega_1)}{\chi_{SSP}^{(2)QB}(\omega_2 + \omega_1, \omega_2, \omega_1)} \approx \frac{|k_z(\omega_1 + \omega_2)| + |k_z(\omega_2)| + |k_z(\omega_1)|}{|k_z(\omega_2 - \omega_1)| + |k_z(\omega_2)| - |k_z(\omega_1)|}$$
$$\approx \left| \frac{k(\omega_2 + \omega_1)}{k(\omega_2 - \omega_1)} \right| \quad (49)$$

One can then simply compare the two DFG responses (DFG and DFG'), as in Eq. 50.

$$\frac{\chi_{SSP}^{(2)QB}(-\omega_2, -\omega_2 - \omega_1, \omega_1)}{\chi_{SSP}^{(2)QB}(-\omega_2 + \omega_1, -\omega_2, \omega_1)} \approx \left| \frac{k(\omega_2 + \omega_1)}{k(\omega_2)} \right|^2 \left| \frac{k(\omega_2 - \omega_1)}{k(\omega_2 + \omega_1)} \right|$$
$$= \left| \frac{k(\omega_2 + \omega_1) k(\omega_2 - \omega_1)}{k(\omega_2)^2} \right|$$
$$\approx 1 - \frac{k(\omega_1)^2}{k(\omega_2)^2}$$
$$\approx 1 \quad (50)$$

Therefore, the bulk quadrupolar response should show almost no dependence on the upconversion frequency, but does show a significant amplitude scaling factor difference between SFG and DFG pathways.

*Experimental SFG and DFG' Comparison*

As discussed above, since the SFG and DFG' pathways include the same three frequencies, the intrinsic dipolar contributions are free from dispersion and must be equal. Therefore, any difference between the responses from these two pathways can only arise from depth contributions or signals beyond the electric dipole approximation. Figure S2 shows the experimental spectra for these two pathways for both $D_2O$ and $H_2O$ where a distinct offset in the real part can be seen, just as for the comparison between SFG and DFG shown in Figure 2 in the main text. As this difference is only present in the real part, the two responses clearly have very similar phases, but distinctly different amplitudes. Therefore, since amplitude differences arising from dipolar depth signals are necessarily accompanied by significant phase differences (as discussed above), the origin of this effect must be from signals generated beyond the electric dipolar approximation.

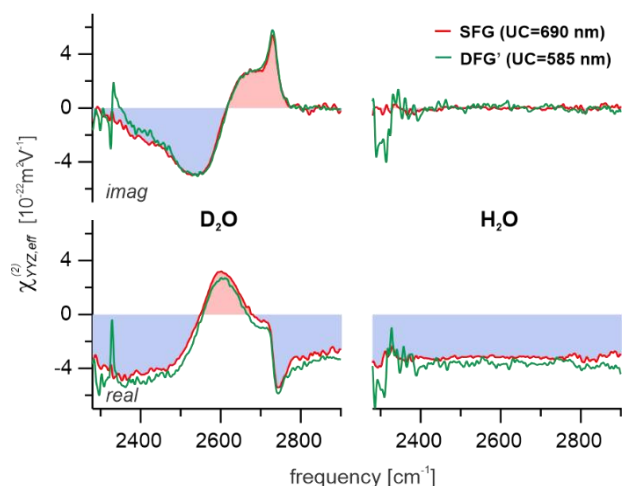

*Figure S2: SFG and DFG' spectra for both $D_2O$ and $H_2O$.*

Based on the theoretical comparison between SFG and DFG' for the different quadrupolar contributions above, this difference likely sources from the IQ or QB contributions. Specifically, as shown in Eqs. 35 and 36, the IQ contribution is, in principle, different between SFG and DFG' based on dispersion in the gradient of the dielectric function across the interface. If this dispersion is significant, however, the comparison between DFG and DFG' shown in Figure 3 in the main text would also be expected to show similar dispersion effects. Therefore, as the DFG and DFG' responses are almost exactly equal, the effect of dielectric dispersion can be neglected and the IQ mechanism discounted as the origin of the observed discrepancy. By contrast, the QB contribution inherently gives a significant amplitude difference between SFG and DFG, disregard of dispersion effects. It is this contribution that is hence concluded to be the source of the observed discrepancy.

*Analysis of the Measured Water Response*

The measured spectra of the non-resonant response from $H_2O$ for both SFG and DFG with an upconversion of 690 nm (i.e., the simultaneous responses) are shown in Figure S3 along with that from the permutation-symmetric DFG' response with an upconversion at 585 nm (i.e., the output SFG frequency). Both real and imaginary parts have been fitted with a constant over the 2450-2750 cm$^{-1}$ frequency range, with the resulting magnitudes of the three responses given in Table S1.

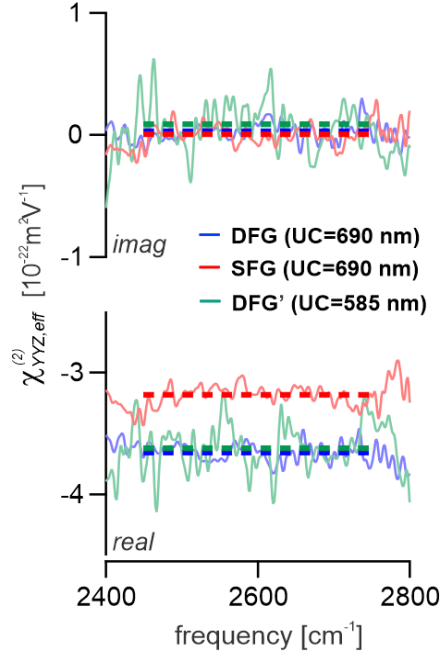

*Figure S3: Real and imaginary parts of the non-resonant $H_2O$ spectra for the SFG, DFG, and DFG' (shifted upconversion) pathways. Also shown are the fitted values for this contribution within the 2450-2750 $cm^{-1}$ frequency range.*

*Table S1: Extracted values for the SFG, DFG, and DFG' susceptibilities from the non-resonant contribution of $H_2O$.*

| Susceptibility (upconversion / nm) | Value / $10^{-22} m^2 V^{-1}$ |
|---|---|
| $\left|\chi_{eff}^{SFG}(690)\right|$ | $3.1842 \pm 0.0046$ |
| $\left|\chi_{eff}^{DFG}(690)\right|$ | $3.6555 \pm 0.0048$ |
| $\left|\chi_{eff}^{DFG}(585)\right|$ | $3.6215 \pm 0.0118$ |

From the data, it is clear that there is a significant measurable difference between the SFG response and either DFG response, but only a minute discrepancy between the two DFG responses. As discussed above, this fits well with the expectation from a bulk quadrupolar contribution to the overall signals.

Using the values for the input frequencies, the theoretical ratio between the bulk quadrupolar contributions to these responses can be calculated from the theory outlined above. For the simultaneously generated responses, this is given in Eq. 51.

$$\frac{\chi_{SSP}^{(2)QB}(-\omega_2+\omega_1,-\omega_2,\omega_1)}{\chi_{SSP}^{(2)QB}(\omega_2+\omega_1,\omega_2,\omega_1)} \approx \left|\frac{k(\omega_2+\omega_1)}{k(\omega_2-\omega_1)}\right| \approx 1.437 \qquad (51)$$

Taking the measured values from Table S1, the ratio of the total responses can then be determined, as in Eq. 52, where the remaining (entirely interfacial) contributions to the overall response are labelled as *I*, and are equal between SFG and DFG.

$$\frac{\chi_{eff}^{DFG}(690)}{\chi_{eff}^{SFG}(690)} = \frac{\chi_{DFG}^{(2)QB}(690)+\chi_{DFG}^{(2)I}(690)}{\chi_{SFG}^{(2)QB}(690)+\chi_{SFG}^{(2)I}(690)} = 1.1480 \pm 0.0022 \qquad (52)$$

Combining these two equations then allows us to extract the ratio between the bulk quadrupolar contribution and the remaining interfacial contribution, as in Eqs. 53 and 54.

$$1.437\chi_{SFG}^{(2)QB}(690) + \chi_{SFG}^{(2)I}(690) = (1.1480 \pm 0.0022)\left(\chi_{SFG}^{(2)QB}(690) + \chi_{SFG}^{(2)I}(690)\right) \quad (53)$$

$$\frac{\chi_{SFG}^{(2)QB}(690)}{\chi_{SFG}^{(2)I}(690)} = 0.512 \pm 0.009 \quad (54)$$

This value of 0.512 thus suggests that the overall SFG response is ~34% bulk quadrupolar, with the corresponding DFG proportion being ~42%.

Whilst in the theory outlined above, we show that the bulk quadrupolar response is mostly independent on upconversion frequency, there is still a slight deviation of this ratio from unity, as is equally observed in the experimental data in Table S1. This allows us to test the conclusions about the bulk quadrupolar contributions to each response by feeding the extracted ratio in Eq. 54 into the theoretical ratio of the two DFG responses and comparing it to the measured value. This comparison is detailed in Table S2 (also shown in Table 1 in the main text) where there is remarkable similarity between the measured and predicted values, with the deviation being well below the uncertainty.

*Table S2: Comparison between measured and predicted values of the ratio of $H_2O$ non-resonant amplitudes for DFG responses measured with upconversion beams at 585 and 690 nm.*

| Susceptibility Ratio | Measured | Predicted |
|---|---|---|
| $\dfrac{\chi_{eff}^{DFG}(585)}{\chi_{eff}^{DFG}(690)}$ | 0.991±0.003 | 0.988±0.010 |

In addition to quantifying the bulk quadrupolar contribution to the non-resonant response, we can also assess its significance to the resonant response. Unlike the non-resonant contribution which is frequency-independent and thus yields a constant value for the individual SFG and DFG spectra and thus also their amplitude ratio, the ratio for the resonant response can be influenced by the vibrational line-shapes of the different source terms, and thus is not generally independent of frequency. Therefore, rather than assessing the quadrupolar contribution by fitting the spectra with a constant, the amplitude ratio between DFG and SFG (shown in Figure S4) must be assessed in terms of its deviation from unity. The spectrum has been fitted with a constant value between 2500-2700 cm$^{-1}$, representing the average ratio between the spectra within this region of high signal-to-noise. From the fitting, the average comes to 1.00, showing an overall lack of any significant deviation from unity, and thus indicating no significant bulk quadrupolar contributions to the resonant response of the OD stretches of water. Nevertheless, an upper bound can be placed on the significance of any quadrupolar contributions by means of the spread of amplitude ratios within this region. The overall average is then 1.00 ± 0.04, which, based on the predicted amplitude ratio for the bulk quadrupolar contributions of 1.437 shown in Eq. 51 above, shows that their significance to the resonant response must be less than 10%.

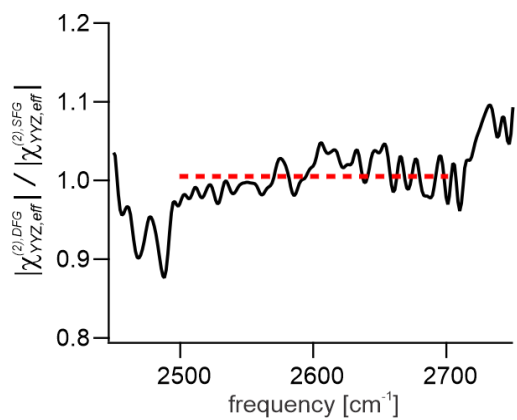

*Figure S4: Amplitude ratio (DFG / SFG) of the purely resonant D$_2$O response, fitted with a constant value.*

### *IQ and IQB Contributions to Resonant and Non-resonant Responses*

In contrast to the non-resonant contribution, it is clear that no amplitude difference between SFG and DFG is observed in the resonant response, and thus any bulk quadrupolar contribution must be negligible. However, as the ID, IQ, and IQB contributions to SFG and DFG are equal, a comparison between their spectra cannot separate their relative contributions. Nevertheless, with the QB term only being significant in the non-resonant response, the quadrupolar mechanisms are suggested to be dominated by interactions with the electronic part of the molecular wavefunction rather than its vibrational component. Therefore, as shown in the theoretical description above, the IQ and IQB mechanisms are also suggested to be insignificant contributions to the resonant response as they arise from the same intrinsic mechanisms i.e., D1, D2, and Q. This suggests that the resonant response is predominantly dipolar in origin and thus reports on the structural anisotropy at the interface.

For the non-resonant response, however, it is not immediately clear how much of the response arises from the ID, IQ, and IQB mechanisms, with them potentially all being significant contributors alongside the QB term. As discussed in the main text, however, the discrepancy between the observed anisotropic depths for the resonant and non-resonant responses (7.7 vs 3.1 Å, respectively) does suggest that they arise from different mechanisms. Whilst this, to some extent, is expected as they probe different parts of the molecular wavefunction: vibrational vs. electronic, with the observation that the anisotropy seems to be predominantly due to the change in H-bonding environment, it is likely that the anisotropy in molecular vibrational structure and the electron cloud are similar. Therefore, the reduced depth of the non-resonant response may well instead be an indicator for interfacial quadrupolar contributions since these are dominated by the anisotropy in the dielectric function at the interface. This conclusion also fits well with previous suggestions by Tahara and co-workers who, through polarisation-dependent measurements, determined that the non-resonant contributions from air-liquid interfaces includes a substantial IQ contribution.(26) We speculate, therefore, that the non-resonant response of water is predominantly quadrupolar, with an interfacial contribution that only depends on the input frequencies through their dielectric functions and a bulk contribution that is almost frequency-independent, but depends on incident angle owing to the z-component of the wave-vector mismatch and the specific pathway being probes (SFG or DFG).

## The Non-resonant Response as a Phase Reference

As discussed in the main text, the non-resonant response from water has been suggested as a possible phase reference for SFG. Despite its exact phase being previously contentious, it is shown here to be very close to ±180°. Specifically, with an apparent depth of 3.1Å, the overall

SFG response has a phase of -179.6°. To make a good phase reference, its phase should be well-known and largely independent of the specific experimental set-up, including the incident angles and input frequencies. While a non-zero depth means the response is dependent on the coherence length and thus on both of the above experimental factors, with anisotropic contributions that decay over well below 1 nm, such dependency is negligible. Furthermore, the bulk quadrupolar response is precisely always 180° in phase. Therefore, the combination of small-depth interfacial terms with the BQ response makes the non-resonant contribution an excellent phase reference. By contrast, the amplitude of the non-resonant response will be both frequency and incident angle dependent owing to the frequency-dependence of the dielectric function and wavevector mismatch, as well as the obvious dependence of the incidence angle on the z-projection of the latter.

## Frequency Dependence of the SFG-DFG Phase Difference

Figure 4(a) in the main text shows the extracted phase difference between the SFG and DFG spectra for both the non-resonant response of $H_2O$ and the purely resonant response of $D_2O$. When assessing the resulting spectra, both responses are fitted with constant values to obtain their average and extract their corresponding anisotropic decay lengths. Whilst this approach is entirely representative of the non-resonant response as this should be independent of frequency, it neglects any possible frequency-dependence of the phase difference for the resonant response. This, for example, could arise from differing decay lengths for the different vibrational resonances corresponding to specific structural motifs. On closer inspection, the resonant phase difference reveals interesting features which could indeed be indicative of the different structural motifs having differing spatial distributions. Specifically, it appears that the free OD species which gives rise to the resonance at 2740 cm$^{-1}$ has a phase difference of ~0°, suggesting it is only present in the top-most layer, in accordance with expectation. In contrast, the hydrogen bonded mode at ~2540 cm$^{-1}$ reports a phase difference of ~1°, suggesting it originates from slightly below the phase boundary. However, when comparing the phase differences for the resonant and non-resonant responses, they both show similar amplitude deviations from their respective mean values. This suggests that the extracted phase difference spectra may contain appreciable distortions (systematic errors). Therefore, for more conclusive insight into such frequency dependence, more experimental work and analysis would be required at even higher precision, which is currently beyond our experimental capabilities.